%% file: main.tex
\documentclass[letterpaper, 11pt]{article}

\input{rishi-header}

\input{commands}

\newcommand{\polylog}{{\rm polylog}}

\title{Efficient Strongly Polynomial Algorithms for Quantile Regression}
\date{}

\usepackage{authblk}
\author[1]{Suraj Shetiya}
\author[2]{Shohedul Hasan}
\author[3]{Abolfazl Asudeh}
\author[4]{Gautam Das}
\affil[1,4]{University of Texas at Arlington}
\affil[2]{Google}
\affil[3]{University of Illinois Chicago}
\begin{document}

\begin{titlingpage}
\maketitle
\begin{abstract}
Linear Regression is a seminal technique in statistics and machine learning, where the objective is to build linear predictive models between a response (i.e., dependent) variable and one or more predictor (i.e., independent) variables. One of the classical and widely used approaches is Ordinary Least Square Regression ({\em OLS}). In this paper, we revisit another classical approach, Quantile Regression ({\em QR}), which is statistically a more robust alternative to {\em OLS}. 
However, while there exist efficient algorithms for {\em OLS}, the state-of-art algorithms for {\em QR} require solving large linear programs using interior point methods which are weakly polynomial.
% However, while {\em OLS} is highly scalable and therefore widely used, historically, {\em QR} has suffered from serious scalability issues, with an unacceptably high computational complexity due to its dependence on solving very large and high dimensional linear programs, requiring resource-intensive interior as well as exterior point methods.

Towards filling this gap, this paper proposes several efficient strongly polynomial algorithms for {\em QR} for various settings. For two dimensional {\em QR} (i.e., one dependent and one independent variable), we make a connection to the geometric concept of $k$-set, and propose an algorithm with a deterministic worst-case time complexity of $\mathcal{O}(n^{4/3}\polylog(n))$ and an expected time complexity of $\mathcal{O}(n^{4/3})$ for the randomized version.  We also propose a randomized divide-and-conquer algorithm - \randQR with an expected time complexity of $\mathcal{O}(n\log^2{(n)})$ for the two dimensional QR problem. For the $d$-dimensional QR problem (i.e., one dependent and $d-1$ independent variables), our \randQR algorithm has an expected time complexity of $\mathcal{O}(dn^{d-1}\log^2{(n)})$.
\end{abstract}
\end{titlingpage}

\input{sections/intro}
\input{sections/prelim}
\input{sections/general}
\input{sections/2d_soln}
\input{sections/zpp_approach}
% \input{sections/discussion}

\section{Final remarks}

In this paper, we studied the Quantile Linear Regression problem in two and higher dimensions.
Our first contribution is the \un which helps us update the optimization function efficiently ($\mathcal{O}(d)$) while navigating the neighboring vertices (intersection points) in dual space.
We propose two strongly polynomial approaches to solve the QR problem.
The $k$-set based approach is a parameter sensitive algorithm that enumerates the $k^{th}$ level of the arrangement while using the \un to compute the optimization functions values.
We prove that its time complexity is better than all known approaches for two dimensions.
The problem of efficient enumeration of $k$-sets in higher dimensions is an open problem which can improve our time complexity our higher dimensions.
Our second approach - \randQR, is a randomized divide-and-conquer approach to the QR problem which splits the $d$ dimensional arrangement into disjoint half-spaces and determining the half-space that contains the optimal solution.
\randQR is also proved to be more efficient than all existing approaches for $2$ and $3$ dimensions including weakly polynomial approaches.
\randQR is faster than all existing strongly-polynomial approaches.
The problem of designing a deterministic algorithm that efficiently finding a hyperplane in $d$ dimensions such that it splits the vertices in \belongR into two ``equal'' parts is an interesting open problem for future work. 

\clearpage
\section*{Appendix}
\appendix
\input{sections/appendix}

\bibliography{references}

\end{document}

%% file: rishi-header.tex
\usepackage{lmodern}  % improves display of text
\usepackage[T1]{fontenc}  % allows LaTeX to output non-ASCII characters better; load after fonts if possible
\usepackage[english]{babel}  % sets language to American English
\usepackage{fullpage}  % 1-inch margins all around
\usepackage{titling}  % can be used to create a title page
\usepackage{caption}  % fixes table--caption spacing in article class
\usepackage{microtype}  % improves typographic details (e.g., kerning)
\usepackage{fnpct}  % adjusts kerning of footnote marks (superscripts) before commas or periods

\usepackage[numbers, sort&compress]{natbib}
\usepackage{graphicx}  % allows inclusion of images in document
\graphicspath{{./images/}}  % location of images in file system

\usepackage{booktabs}
\usepackage{mathtools, amssymb, amsthm}  % mathtools loads amsmath

\allowdisplaybreaks[1]  % allows page breaks within amsmath environments; optional value from 1-4 sets permissibility (4 being the highest)

\renewcommand*{\emptyset}{\varnothing}  % prettier empty set symbol
\usepackage{bm}  % use \bm for bold math symbols

\DeclarePairedDelimiterX\set[1]\lbrace\rbrace{\,#1\,}
\usepackage{tikz}
\usetikzlibrary{babel, positioning}

\usepackage[colorlinks, allcolors=blue]{hyperref}  % enable hyperlinks; load before `ellipsis`
\hypersetup{
    pdfinfo={
        Title={Online Maximum Independent Set of Hyperrectangles},
        Author={Anonymous author(s)},  % TODO update for arXiv submission
        Keywords={maximum independent set; hyperrectangle; geometric intersection graph; online algorithm; competitive ratio; adversary model}
    }
}
\usepackage{ellipsis}  % adjusts spacing of ellipses before punctuation

%% file: commands.tex
\usepackage{xspace}
\usepackage{enumitem}
\usepackage{tcolorbox}
\usepackage{multirow}
\usepackage{amsfonts}
\usepackage{amsopn}
\usepackage{graphicx}
\usepackage{algorithm}
\usepackage{fontawesome}
\usepackage{algorithmic}
\usepackage{wrapfig}

\newtheorem{theorem}{Theorem}
\newtheorem{lemma}{Lemma}

\newcommand{\eop}{\hspace*{\fill}\mbox{$\Box$}}     % End of proof
\newcounter{example}%[section]
\renewcommand{\theexample}{\arabic{example}}

\newcommand\stitle[1]{\vspace{2mm}\noindent{\bf #1}\xspace}

\newcommand{\un}{{\sc UpdateNeighbor} subroutine\xspace}
\newcommand{\twodAlg}{{\sc QReg2D}\xspace}
\newcommand{\neighborBaseline}{{\sc QR2D-Neighbor-Baseline}\xspace}
\newcommand{\ksetBaseline}{{\sc QR2D-k-Set-Baseline}\xspace}

\newcommand{\randalg}{Randomized Incremental algorithm}

\newcommand{\SplitOrc}{{\sc Split} function\xspace}
\newcommand{\SplitFn}{{\sc Split}}
\newcommand\ConsSamplFn{{\sc ConstrainedSampling}}
\newcommand\highlight[1]{\textcolor{black}{#1}}
\newcommand\randQR{{\sc RandomizedQR}\xspace}
\newcommand\ConsSampl{{\sc ConstrainedSampling}\xspace}
\newcommand{\belongR}{embraced vertices }

%% file: sections/intro.tex
\section{Introduction}
Linear Regression is a seminal technique in statistics and machine learning, where the objective is to build linear predictive models between a response (i.e., dependent) variable and one or more predictor (i.e., independent) variables from a given dataset of $n$ instances, where each instance is a set of values of the independent variables and the corresponding value of the dependent variable. One of the classical and widely used approaches is {\em Ordinary Least Square Regression} ({\em OLS}), where the objective is the minimize the average squared error between the predicted and actual value of the dependent variable. Another classical approach is {\em Quantile Regression} ({\em QR}), where the objective is to minimize the average weighted absolute error between the predicted and actual value of the dependent variable. {\em QR} (also known as ``Median Regression'' for the special case of the middle quantile), is less affected by outliers and thus statistically a more robust alternative to {\em OLS}~\cite{john2009quantile,haupt2014quantile}. 
% \textcolor{red}{Gautam: all these above need references.}
However, while there exist efficient algorithms for {\em OLS}, the state-of-art algorithms for {\em QR} require solving large linear programs with many variables and constraints. They can be solved using using interior point methods~\cite{portnoy1997gaussian} which are weakly polynomial (i.e., in the  arithmetic computation model the running time is polynomial in the number of bits required to represent the rational numbers in the input), or using Simplex-based exterior point methods which can have exponential time complexity in the worst case~\cite{deza2008good}. 
% \textcolor{red}{Gautam: all the above computational complexity statements need references.}

The main focus of our paper is an investigation of the  computational complexity of Quantile Regression, and in particular, to design efficient strongly polynomial algorithms (i.e., in the arithmetic computation model the running time is polynomial in the number of rational numbers in the input) for various special cases of the problem. 

\subsection{Prior Work} 
In prior work, 
%the two prominent classes of techniques to solve the  Quantile Regression problem are Simplex-based~\cite{barrodale1973improved} and interior-point based~\cite{portnoy1997gaussian} approaches.
Barrodale and Robert ({\em BR}) proposed a Simplex-based  exterior point technique~\cite{barrodale1973improved}, which moves from one exterior point (i.e., a ``corner'' of the feasible polytope defined by the linear program) to another exterior point, in the direction of steepest gradient descent. 
The core idea of {\em BR} originates from Edgeworth's bi-variate weighted median based approach \cite{edgeworth1888xxii}. For the special case of the 2-dimensional QR problem (i.e., one dependent variable and one independent variable), the time complexity of this algorithm is  $\mathcal{O}(n^2)$, which is still the best known strongly polynomial result for 2-dimensions. For the general $d$-dimensional QR problem (i.e., one dependent variable and $d-1$ independent variables), the best known strongly polynomial algorithm is a naive baseline method that runs in $\mathcal{O}(d\, n^{d+1})$ time (note that this assumes the dimension $d$ is bounded).
% All the exterior points based approaches are only suitable for small problem instances with around $1000$ points and few attributes \cite{portnoy1997gaussian}.
Later, Portnoy and Koenkar  proposed the interior point method ({\em IPM})  which finds the optimal solution by minimizing the difference between primal and dual objective cost~\cite{portnoy1997gaussian}.
This method is reasonably fast for larger input sizes in practice, but the worst-case theoretical time complexity is $\mathcal{O}(n^{2.5}\log {1/\epsilon})$ where $\epsilon$ is the desired accuracy, which needs to be set to $\mathcal{O}(2^{-L})$ to obtain the optimal result, where $L$ is the number of bits in the input instance \cite{wright1997primal}. Thus, while the worst case time complexity is independent of the dimension of the QR problem, it is weakly polynomial. Moreover, as shall be clear in later sections,  even for a low-dimensional QR problem, e.g., $d=2$, the corresponding LP formulation can be extremely high-dimensional, since it requires the addition of $\mathcal{O}(n)$ variables. 
% \textcolor{red}{Gautam: check the preceding sentence.} 
Thus, one cannot leverage well-known efficient and strongly polynomial low-dimensional LP algorithms for solving low dimensional QR problems.

\begin{wraptable}{r}{10cm}
%\vspace{-5mm}
% \vspace{-2mm}
\begin{tabular}{|@{}c@{}|l@{}|l@{}|}
\hline
\:Dimension\: & Weakly polynomial & Strongly Polynomial \\ \hline
$d = 2$ & LP - $\Tilde{\mathcal{O}}(n^{2+1/18}L)$ \cite{jiang2020faster}  & $\mathcal{O}(n^2)$\cite{barrodale1973improved} \\ \cline{1-1} \cline{3-3}
General $d$ & LP - $\mathcal{O}(n^{5/2}L)$~\cite{vaidya1989speeding} & Baseline - $\mathcal{O}(d\, n^{d+1})$ \\
\hline                         
\end{tabular}
\vspace{-3mm}
\caption{A summary of results from prior work}
\vspace{-3mm}
\label{tab:summary-prior-work}
\end{wraptable}

Table~\ref{tab:summary-prior-work} provides a snapshot of the computational complexity of these prior techniques.
Besides these  results, there have been prior research on QR algorithms under various distributional assumptions of the input data, which are less relevant to the focus of our paper. 
We discuss the challenges and some of the state of the art techniques in Appendix~\ref{subsec:qt-challenges-state-of-art}.

\subsection{Our Technical Contributions} 
In this paper, we make the following key technical contributions that give rise to several strongly polynomial algorithms.

Our first contribution is to capitalize on the computational geometry concepts of {\em arrangement} and {\em duality}~\cite{edelsbrunner1987algorithms}, and map {\em QR} via a duality transform into a problem of
traversing an arrangement of hyperplanes in search of the intersection point that optimizes the {\em QR} objective function. 
To aid this traversal, we leverage the specifics of the {\em QR} objective function and design an algorithm named \un, which can  update the objective function calculations from a neighboring point in the arrangement very efficiently with $\mathcal{O}(d)$ time and $\mathcal{O}(d)$ space complexity (i.e., each update is independent of $n$). 
This is done by maintaining various aggregate information along every feature.
% with both positive and negative errors.
% By using \un, we develop our quantile regression algorithm in 2-dimension.
This \un  allows us to easily improve the naive baseline QR algorithm's running time from $\mathcal{O}(d\, n^{d+1})$ to $\mathcal{O}(dn^d)$. 
% \textcolor{red}{Gautam: check this. Also, should we give this a name, and add it in our results Table 2 as part of our deterministic result for any d?}  
% \suraj{I have added it to the table. Can we name it \un in higher dimension ?}

Our second contribution is to connect the QR problem with the concept of {\em geometric $k$-sets} ~\cite{edelsbrunner1987algorithms}, which allows our algorithm to restrict its traversal to within the $k^{th}$ level of the arrangement. In our case, the parameter $k$ is set to the quantile, e.g., for median regression, $k = n/2$.
This connection with $k$-sets gives rise to an efficient quantile regression algorithm for 2-dimension, named \twodAlg.
The deterministic time complexity for our {\em QR} algorithm is $\mathcal{O}(nk^{1/3}\log^{1+a}{(n)})$ where  $a>0$ is an arbitrarily small constant.
If we use probabilistic $k$-set enumeration procedures in our algorithm, the expected time complexity of our algorithm is $\mathcal{O}(n^k{1/3})$. {\em These are asymptotically better than any other existing  exterior or interior point approaches in two dimensions} (recall that {\em BR} runs in $\mathcal{O}(n^2)$ time while {\em IPM} runs in $\Tilde{\mathcal{O}}(n^{2.5}\log{(1/\epsilon)})$ time).
% \suraj{Should we discuss how $k$-set idea fares with changes in  $d$?}
% \textcolor{red}{Gautam: we should. check what I have written below.}
In higher dimensions, counting as well as efficiently enumerating $k$-sets are generally challenging and open problems in computational geometry. Any breakthroughs are likely to have implications in our $k$-set based algorithms for higher dimensional QR problems.

Our third contribution is a probabilistic \randQR algorithm for the QR problem in general $d$ dimensions.
We devise a divide-and-conquer approach to split the $d$ dimensional arrangement into two half-spaces based on a randomly chosen hyperplane and determine the half-space that contains the optimal solution. The highlights of our approach include developing an efficient probabilistic technique for sampling uniformly at random vertices contrained within a portion of the arrangement, which we further connect to the problem of counting inversions of a permutation.
Our \randQR algorithm has an expected time complexity of  $\mathcal{O}(n\log^2{(n)})$ for two dimensions which is asymptotically better than the other strongly and weakly polynomial algorithms.
The time complexity of \randQR in higher dimensions is $\mathcal{O}(d\, n^{d-1}\log^2{(n)})$, which is faster than the known deterministic strongly polynomial algorithms.

A summary of the time complexities of our algorithms is presented in Table~\ref{tab:summary}.

% \stitle{Paper Organization:}
% In Section \ref{sec:preliminaries}, we introduce the problem with all the necessary details to understand the problem and solutions in this paper. 
% In Section \ref{sec:geom-map}, we describe the geometric mapping of {\em QR} problem and related definitions and important {\em QR} properties.
% We present our \un algorithm in Section \ref{sec:general}, the  quantile regression algorithm in $2$-dimensions \twodAlg  in \ref{sec:2d}, challenges in higher dimension in \ref{sec:discussion} and followed by related work Section.
\begin{table}[]
\centering
% \caption{A summary of our algorithmic results; our contribution highlighted in \textcolor{blue}{blue} and by * marker.}

\begin{tabular}{|@{}c@{}|l@{}|l@{}|}
\hline
% \multicolumn{1}{l|}{Dimension} & Strongly Polynomial & Comments \\ \hline
Dimension & Strongly Polynomial & Comments \\ \hline
\multirow{3}{*}{$d=2$} & \highlight{$\mathcal{O}(nk^{\frac{1}{3}}\log^{1+a}{(n)})$} - \twodAlg~\ref{alg:2d-algo} & Lemma~\ref{thm:qrreg2d-proof} \\ \cline{2-3} 
    & Expected \highlight{$\mathcal{O}(nk^{\frac{1}{3}})$} - \twodAlg~\ref{alg:2d-algo} & Lemma~\ref{thm:qrreg2d-rand-proof} \\ \cline{2-3} 
    & Expected \highlight{$\mathcal{O}(n\log^{2}{(n)})$} - \randQR~\ref{alg:generic-zpp-algo}& Theorem~\ref{thm:expected-runtime-2d-zpp} \\ \hline
$d=3$ &  Expected \highlight{$\mathcal{O}(n^2\log^{2}{(n)})$} - \randQR~\ref{alg:generic-zpp-algo} & Theorem~\ref{thm:expected-runtime-3d-zpp} \\
% \multirow{2}{*}{3} & *: Expected \highlight{$\mathcal{O}(n\log^{2}{(n)}L)$} \ & *: Expected \highlight{$\mathcal{O}(n^2\log^{2}{(n)})$} - \\ 
 % &  \ \ Theorem~\ref{thm:expected-runtime-3d-zpp-weakly-polynomial} & \ \  Theorem~\ref{thm:expected-runtime-3d-zpp} \\
\hline
\multirow{2}{*}{General $d$}    & Expected \highlight{$\mathcal{O}(d\, n^{d-1}\log^{2}{(n)})$} - \randQR~\ref{alg:generic-zpp-algo} & Theorem~\ref{thm:expected-runtime-3d-zpp} \\
& $\mathcal{O}(dn^d)$ - \un & Theorem~\ref{thm:constant_update}\\
    % \cline{2-2} 
    % &$\mathcal{O}_p(n^{1+\alpha}d\log{(n)})$~\cite{sonnevend1991complexity} & \\ %\hline
             \hline                    
\end{tabular}
\vspace{-2mm}
\caption{A summary of our algorithmic results.}
\label{tab:summary}

\end{table}

%% file: sections/prelim.tex
\section{Preliminaries}\label{sec:preliminaries}

% \shohed{
%     \begin{itemize}
%         \item Notations
%         \item Problem Definition
%     \end{itemize}
% }

In this section, we provide useful definitions and notations, as well as a formal description of the main problem considered in this paper. Some of the notations and formal problem definitions are borrowed from existing literature such as \cite{barrodale1973improved,portnoy1997gaussian}. 

\subsection{Running Example}
% \begin{table}[t]
% \centering
% \begin{tabular}{|c | c |c |} 
%  \hline
%  $id$ & $A_1$ & $A_2$ \\ %[0.5ex] 
%  \hline
%  $t_1$ & 3.15 & 3.13 \\ 
%  $t_2$ & 1.97 & 1 \\
%  $t_3$ & 1.369 & 2.43 \\
%  $t_4$ & 0.149 & 1.287\\
%  $t_5$ & -0.39 & 0.222 \\ 
%  $t_6$ & -0.51 & -0.65\\
%  $t_7$ & -2.04 & 7.30 \\ [1ex] 
%  \hline
% \end{tabular}
% \caption{A 2D dataset.}
% \vspace{-5mm}
% \label{tbl:running_ex}
% \end{table}

\begin{wraptable}{r}{3.3cm}
%\vspace{-5mm}
\vspace{-20mm}
\begin{tabular}{|c | c |c |} 
 \hline
$id$ & $A_1$ & $A_2$ \\ %[0.5ex] 
 \hline
 $t_1$ & 3.15 & 3.13 \\ 
 $t_2$ & 1.97 & 1 \\
 $t_3$ & 1.369 & 2.43 \\
 $t_4$ & 0.149 & 1.287\\
 $t_5$ & -0.39 & 0.222 \\ 
 $t_6$ & -0.51 & -0.65\\
 $t_7$ & -2.04 & 7.30 \\
%  \hline
 \hline
\end{tabular}
\caption{A 2D dataset.}
\vspace{-4mm}
\label{tbl:running_ex}
\end{wraptable}

% \begin{table}[!ht]
% \centering
% \begin{tabular}{|c | c |c |c | c |c |c | c |} 
%  \hline
%  $id$ & $t_1$ & $t_2$ & $t_3$ & $t_4$ & $t_5$ & $t_6$ & $t_7$ \\
%  \hline
%  $A_1$ & 3.15 & 1.97 & 1.369 & 0.149 & -0.39 & -0.51 & -2.04 \\
%  $A_2$ &  3.13 & 1 & 2.43 & 1.287 & 0.222 & -0.65 & 7.30 \\
%  \hline
% \end{tabular}
% \caption{A 2D dataset.}
% \vspace{-2mm}
% \label{tbl:running_ex}
% \end{table}

In this paper, we use a toy dataset, shown in Table \ref{tbl:running_ex}, which will be used throughout the paper.
% \textcolor{red}{Gautam: have Table 3 right here, and not floating at the top of the page. I.e., make it appear as a part of section 2, and separate from Table 2}
%The dataset contains seven input points and two attributes $A_1$ and $A_2$.
%Our goal is to predict one of the attributes from the other attribute.
Assume that we want to build a linear model to predict $A_2$ from $A_1$.
%For simplicity, let us denote $A_1$ and $A_2$ by $x$ and $y$, respectively.
Figure \ref{fig:input_space} shows a scatter plot of the data points in 2-dimensional space (where the $x$-axis is $A_1$ and $y$-axis is $A_2$), and two linear models, one being the classical {\em OLS} model and the other being the classical {\em QR} model. The former minimizes the sum of squared errors, while the latter minimizes the sum of absolute errors (this is also known as  {\em $\ell_1$-regression}), where the error is defined as the difference between the actual and predicted value of the $A_2$ variable. 
% where each input tuple is a point with $(x,y)$ coordinate.
%We get a line if we fit a {\em Linear Regression (LR)} model for these points. 
%Figure \ref{fig:input_space} demonstrates two classical fitted {\em LR} models. 
%One of the {\em LR} models is {\em Ordinary Least Square Regression(OLS)}, which minimizes the sum of squared error.
%The other {\em LR} model is a special type of {\em Quantile Regression(QR)} ({\em $\ell_1$-regression}), which minimizes the sum of absolute error.
%For any input point, the vertical distance between the point and the {\em LR} model is the error in regression.
%In  Figure \ref{fig:input_space}, the red and green dotted lines show the error for  {\em OLS} and {\em $\ell_1$-regression} models, respectively.
The figure shows that the {\em OLS} model is heavily affected by the presence of outlier $t_7$, however, the {\em $\ell_1$-regression}  is less sensitive to the outlier.
%For all but one point, the length of a green dotted line is smaller than that of a red dotted line.
% \begin{figure}[tp]
%     \centering
%     % \includegraphics[width=0.45\textwidth]{figures/InputDiagram.PNG}
%     \includegraphics[width=0.31\textwidth]{figures/InputSpace2.PNG}
%     \caption{The Visual demonstration of database}
%     \vspace{-5mm}
%     \label{fig:input_space}
% \end{figure}

\subsection{Quantile Regression Problem Definition}

% % {\em $\ell_1$ norm, i.e., sum of the absolute error.}

% In linear regression, we fit a linear model for the given dataset. 
% In 2-dimension a linear fit is a line.
% The equation of a line is $\hat{y}=\beta_0+\beta_1x$ where $\beta_0$ is the intercept, $\beta_1$ is slope and \hat{y} is regressed value. 
% For 2-dimension, the relationship between y and x can be expressed as 
% For example,
% in  Figure \ref{fig:input_space}, we fit two different linear models for our data.
% One of the linear model is a specific type of Quantile Regression known as {\em $\ell_1$} regression. 
% The other linear model is Ordinary least Square Regression (OLS).
% From Figure \ref{fig:input_space}, OLS is heavily affected by the outlier $t_7$. 

% The Quantile Regression is a
% In 2-dimension a linear fit is a line.

% Let us denote $A2$ by $x$ and A1 by y for the simplicity. 

% Our goal is to fit a linear regression for this dataset. 

\stitle{Dataset:} 
Let $\mathcal{D}$ be a dataset with $n$ points and $d$ numerical variables. 
%Our goal is to predict one of the attributes from the rest of the attributes. 
%The attribute we predict is called \textit{response variable}.
%The attributes used in the prediction are called \textit{predictor variables}.
%For our running example, $y$ ($A_2$) is the response variable and $x$ ($A_1$) is the predictor variable.
We denote the response variable and the vector of predictor variables by  $y_i$ and $X_i=[1\ x_1\ x_2 \ldots x_{d-1}]$  respectively for $i^{th}$ point.
A constant $1$  is added in $X_i$ to make the problem definition (defined later) consistent.
% For $d$ attributes, we denote the response variable and the vector of predictor variables by  $y_i$ and $X_i=[1\ x_1\ x_2 \ldots x_{d-1}]$  respectively for $i^{th}$ point.
% A constant $1$  is added in $X_i$ to make the problem definition (defined later) consistent.
$X$ is a matrix where $i^{th}$ row corresponds to $X_i$.  
$X$ and $y$ for the running example are shown in Figure~\ref{tbl:running-example-2d}.
% \vspace{-2mm}

% \begin{align*}
% /[]
% X^T&=\begin{bmatrix}
% 1 \ \ & 1 \ \ & 1 \ \  & 1 \ \  & 1 \ \  &1\ \ \  & 1 \\
% 50\ \ \ & 24\ \  & 12\ \  & 5.50  & 2.40 &1\ \ \  & 0.05
% \end{bmatrix}\\
% % \]
% % \[
% Y^T &= \begin{bmatrix}0.60 & 0.55 & 0.45 & 0.35 & 0.28 & 0.20 & 0.12 \end{bmatrix}
% % \]
% \end{align*}
% \vspace{-3mm}
% \[
% X=
%   \begin{bmatrix}
%    1 & 50\\ 
%  1 & 24\\
%  1 & 12\\
%  1 & 5.5\\
%  1 & 2.4\\ 
%  1 & 1 \\
%  1 & 0.05 
%   \end{bmatrix}
% , and\ \ y = 
%   \begin{bmatrix}
%    0.6\\ 
%  0.55\\
%  0.45\\
%  0.35\\
%  0.28\\ 
%  0.2 \\
%  0.12 
%   \end{bmatrix} 
% \]
\begin{wrapfigure}{r}{0.61\textwidth}
    \centering
    \vspace{-10mm}
    \begin{align*}
    % /[]
    X^T&=\begin{bmatrix}
    1 \ & 1 \ & 1 \ & 1 \ & 1 \ &1\  & 1 \\
    3.15\ & 1.97  & 1.369 & 0.149  & -0.39 & -0.51  & -2.04
    \end{bmatrix}\\
    % \]
    % \[
    Y^T &= \begin{bmatrix}3.13 & \ \ 1\ \ \ &\ 2.43 & 1.287 &\ 0.222 &\ -0.65 &\ \ 7.30 \end{bmatrix}
    % \]
    \end{align*}
    \vspace{-6mm}
    \caption{Running example in two dimensions}
    \label{tbl:running-example-2d}
    \vspace{-4mm}
\end{wrapfigure}

\begin{figure*}[t] 
\centering
    \begin{minipage}[t]{0.48\linewidth}
        \centering
        \includegraphics[width=0.98\textwidth]{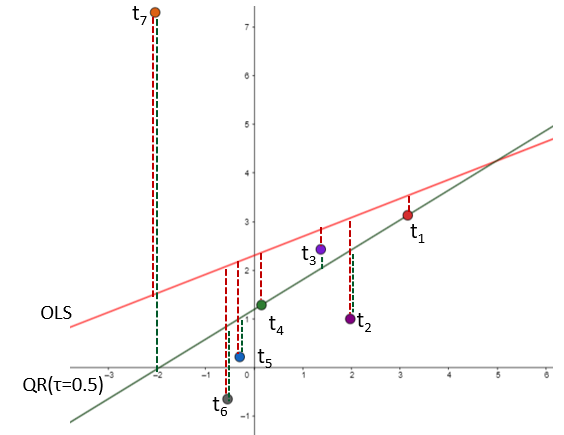}
    \caption{The (primal space) visualization of the running example dataset}
    \label{fig:input_space}
    \end{minipage}
    \hfill
    \begin{minipage}[t]{0.48\linewidth}
        \centering
        \includegraphics[width=0.98\textwidth]{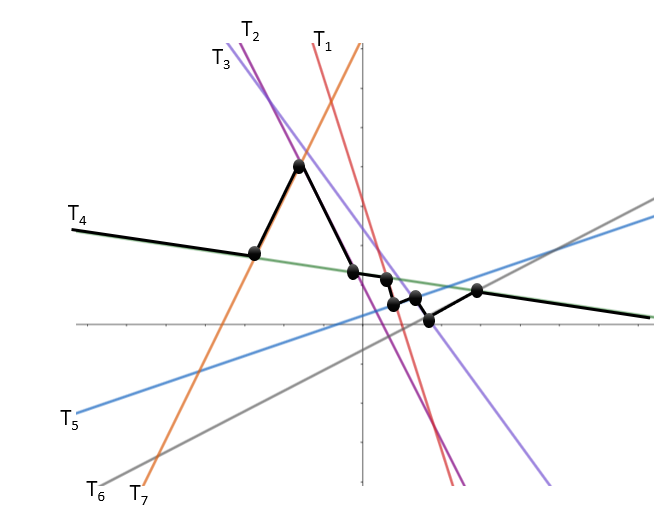}
        \caption{The dual space visualization of the running example dataset, with the 4-th level of its arrangement highlighted}
        \label{fig:dual_space}
    \end{minipage}
\end{figure*}

\stitle{Residual:} If the actual value is $y$ and the predicted value is $\hat{y}$, then the residual, $r$, can be defined as $y-\hat{y}$.
% \textcolor{red}{Gautam: what are these terms``vector of actual values", ``vector of regressed values''? Not defined.}
The value of $r$ can be positive, negative, or zero.
From this viewpoint, we can also define  $r_i = r_i^{+} -r_i^{-}$ where,  for each point $i$ in $\mathcal{D}$, $r^{+}_i = max(r_i,0)$ and $r^{-}_i = -min(r_i,0)$. 
% Note that  $r_i^+$, and $r_i^-$  contain only non-negative values. 
The dotted lines in Figure \ref{fig:input_space} show the residuals. 
$t_5$ has positive residual and $t_7$ has negative residual.

\stitle{Quantile Parameter, $\tau$:} In Quantile Regression, the quantile parameter $\tau \in (0, 1)$ determines what fraction of a total number of input points will have a negative residual.
For $\tau = 0.5$, half of the input point will have negative residuals, and the remaining half will have positive residuals.
% In practice, median, quartiles, deciles, 0.05, and 0.95 are some of the most frequently used quantile parameters.

% \stitle{Quantile Parameter, $\tau$:} In Quantile Regression, quantile parameter $\tau \in (0, 1)$ determines how the regression hyperplane divides the data points.
% % \textcolor{red}{Gautam: what is regression hyperplane? It is not defined. You cannot have a preliminary section written where notations are being introduced without being defined first.}
% For $\tau = 0.5$, the regression hyperplane divides the data points into two equal parts which is also known as median quantile regression.
% In practice, median, quartiles, deciles, 0.05, and 0.95 are some of the most  frequently used quantile parameters.

% \textbf{Regression Parameter, $\beta$:} Quantile regression  finds a parameter vector $\beta=\{ {\beta}_0, {\beta}_1, \ldots, {\beta}_{d-1}\}$ which represents the best regression hyperplane.
% Here, ${\beta}_0$ is regression hyperplane intercept and other parameters represent slope along the axis of predictor features. 
% Let $\hat{y}$ be a vector of predicted target values.
% For a given $\beta$ and $i^{th}$ point, the estimated target value is $\hat{y_i} = X_i^T\beta$. 
% For Table \ref{tbl:running_ex}, if we are given  $\beta=[0.5\ 0.2]$, and i =1, the  $\hat{y_1}$ becomes $0.5\times1 + 0.2\times 0.85 = 0.67$. 

\stitle{Formal Problem Definition:} 
% Quantile Regression  estimates the conditional quantile value of a response variable from the linear combination of predictor variables. 
% \textcolor{red}{Gautam: what is ``dependent variable''? I thought you had defined something called ``response variable'' earlier.} 
The formal definition of {\em Quantile Linear Regression (QR)}  problem in this paper is defined as follows. 

\begin{tcolorbox}
% \begin{center}\textbf{Quantile Linear Regression}
% \end{center}
\stitle{Quantile Linear Regression}: 
Given the quantile parameter $\tau$ and a dataset $\mathcal{D}$ with $n$ points, where $y$ is a vector of the actual response variable and $X$ is a $n\times d$ matrix constructed from predictor variables, find a parameter vector $\beta \in \mathbb{R}^d$ that optimizes the following objective function.
\vspace{-6mm}
% \begin{align*}
% \underset{\beta\in \mathbb{R}^d}{\min}\ \overset{n}{\underset{i=1}{\sum}}&\ \tau r^+_i\ +\ (1-\tau)  r^-_i \\ &where \ r^+_i =  max(y_i - X_i^T\beta,\ 0)\ \\
% &and\  r^-_i =  -min(y_i-X_i^T\beta, 0)
% \end{align*}
\begin{center}
    $\underset{\beta\in \mathbb{R}^d}{\min}\ \overset{n}{\underset{i=1}{\sum}}\ \tau r^+_i\ +\ (1-\tau)  r^-_i$ \\
    s.t. $~r^+_i =  \max(y_i - X_i^T\beta,\ 0)$ \\
    \hspace{10mm}$r^-_i =  -min(y_i-X_i^T\beta, 0)$
\end{center}
\end{tcolorbox}

% \begin{align*}
%     \underset{\beta\in \mathbb{R}^d}{\min}\ \underset{i\in \{i:y_i \geq   X_i\beta\}}{\sum}&\tau\ (y_i - X_i\beta)\ \\
%     +&\ \underset{i\in \{i:y_i < X_i\beta\}}{\sum}(1-\tau)\ (X_i\beta - y_i) \\
% where r^+_i =  max(y_i - X_i\beta,0) \\ 
% \ and\  r^-_i =  -min(y_i-X_i\beta,0)

% \textcolor{red}{Gautam: compress the above box into fewer lines}
In Figure \ref{fig:input_space}, we used $\tau=0.5$ for the {\em QR}  which puts equal penalty for both positive and negative residuals.
For $\tau=0.5$, {\em QR} is known as {\em $\ell_1$-regression} (also known as {\em median regression}). Different applications may require different values of $\tau$. For example, if $\tau=0.95$, the objective function puts a high penalty (0.95) on any point with positive residual and a low penalty (0.05) on any point with negative residual.

% In Figure \ref{fig:input_space}, we show a comparison between  {\em $\ell_1$-regression} and {\em OLS}, two classical LR techniques.

% For $\tau=0.5$, the quantile regression problem is known as median regression or, $l_1$ regression problem which is an alternative to OLS. 
% In 
% In conformal prediction, $\tau=0.05$ and $\tau =0.95$ are used to calculate the prediction interval.  
% \textcolor{red}{Gautam: What are these terms: ``OLS", ``conformal prediction"? Not defined.}

\medskip
% \textcolor{red}{Gautam: I am worried that most sigmod readers would be lost by now. This section so far has too many heavy notations, and not enough intuition or running examples. Here is a suggestion. Right in the beginning, even before getting into notations, first introduce a running example. For the example dataset in Table 1, can you have a diagram that shows the solution of the quantile regression line, and shows how it divided the data into two equal parts, and shows dotted vertical lines from each data point to the regression line, and a statement that the sum of the lengths of these dotted lines is being minimized? At the same time, you can also show the OLS regression line, and how it differs from the quantile regression line? That way you can intuitively introduce both types of regression. }

One of the common techniques that has been used to solve the QR problem is Linear Programming.
In the LP formulation, $2n$ variables are introduced, one for each of the residuals.
Additionally, $d$ variables are needed for the optimal QR hyperplane parameters $\beta$.
% Therefore, the LP has a total of $2n+d$ variables.
The LP formulation for the QR problem is given in Figure~\ref{fig:QRLP}:
% \vspace{-5mm}
% \begin{align*}
%     \min \sum_{i=1}^n ( \tau r_i^+ + (1-\tau) r_i^-)
% \end{align*}
% \vspace{-5mm}
% \begin{table}[!h]
% \centering
% \begin{tabular}{lc}
% \multirow{2}{*}{subject to\ \ \ \ \ } & $r_i^+\geq y_i - \sum_{j=1}^d X_i^T\beta$ \ \ \ and \ \ \ $r_i^+\geq 0$ for each $1\leq i \leq n$\\
%     & $r_i^-\geq \sum_{j=1}^d X_i^T\beta - y_i$   \ \ \ and \ \ \ $r_i^-\geq 0$ for each $1\leq i \leq n$\\
% \end{tabular}
% \end{table}

\begin{wrapfigure}{r}{0.31\textwidth}
    % \centering
    \vspace{-9mm}
    \hspace{-1mm}\framebox[0.32\textwidth]{
        \parbox{0.3\textwidth}{
        \small
        \begin{align*}
            ~\mbox{min}&\mbox{imize}~ \sum_{i=1}^n ( \tau r_i^+ + (1-\tau) r_i^-)\\
            \mbox{sub}&\mbox{ject to} &~\\
            &r_i^+\geq y_i - X_i^T\beta\\
            &r_i^-\geq X_i^T\beta - y_i\\
            &r_i^+\geq 0,~ \forall 1\leq i \leq n\\
            &r_i^-\geq 0,~ \forall 1\leq i \leq n %\\
            % &\r_i^-\geq 0,~ \forall 1\leq i \leq n\\
        \end{align*}}
    }
        \vspace{-3mm}
    \caption{QR LP formulation.}
    \label{fig:QRLP}
    \vspace{-14mm}
\end{wrapfigure}

Based on the LP formulation, even if we assume $d$ is bounded, there are $2n+d=\mathcal{O}(n)$ variables and $2n=\mathcal{O}(n)$ constraints.
As the number of variables translate to dimensions in LP and is not a fixed number, the QR problem cannot use some of the strongly polynomial LP results for fixed dimensions.
Table~\ref{tab:summary-prior-work} summarizes some of latest results for the weakly polynomial LP techniques when applied to QR problem.
% \textcolor{red}{Gautam: where is the mapping of QR to a linear program? I think we should give it here.}
A detailed description of the current state of the art algorithms for {\em QR} is presented in Appendix~\ref{subsec:state-of-art}.

% \begin{figure*}[t] 
% \centering
%     \begin{minipage}[t]{0.33\linewidth}
%         	\centering
%         	\includegraphics[width=0.95\textwidth]{figures/InputSpace2.PNG}
%         	% \vspace{-6mm}
%         	\caption{The Visual demonstration of the data set}
%         	\label{fig:input_space}
%     \end{minipage}
%     \hfill  
%     \begin{minipage}[t]{0.32\linewidth}
%         % \centering
%         % 	\includegraphics[width=0.95\textwidth]{figures/DualSpace2.png}
%         % \vspace{-2mm}
%         % \caption{Dual space of the Sample Data set}
%         % \label{fig:dual_space}

%         \centering
%         \includegraphics[width=0.95\textwidth]{figures/QRREG2D.png}
%         % \vspace{-6mm}
%         \caption{Traversing k-level of arrangement in order}
%         \label{fig:qrregseq}
%     \end{minipage} 
%     \hfill
%     \begin{minipage}[t]{0.32\linewidth}
%         	\centering
%         	\includegraphics[width =.95\textwidth]{figures/OracleD.PNG}
%         	% \vspace{-6mm}
%         	\caption{Update Operation}
%         	\label{fig:orderd_update}
%     \end{minipage}
%     % \vspace{-4mm}
% \end{figure*}

\subsection{Other Useful Notations}\label{sec:geom-map}

% Each input point $\mathcal{D}_i$ can be viewed as a point in $d$ dimensional space.
In this section, we define the {\em dual space} of $\mathcal{D}$, its connection with the quantile regression problem, and some of the basic properties and definitions  that we will use throughout the paper.

% \textcolor{red}{Gautam: we need a good definition of the ``quantile regression hyperplane'' well highlighted somewhere.}
% A geometrical view of this division of points into two partitions is know as a $k$-set.

\stitle{{\em QR} Hyperplane, $\mathcal{H}$:} {\em LR} estimates the response variable as a linear combination of predictor variables with an offset. 
% There is a geometric interpretation of an {\em LR} model.
% \textcolor{red}{Gautam: I notice in several places you have mentioned the word ``fit''. What does it formally mean?}
% If we assume the input tuple as points in space, the LR fit is a hyperplane in space.
The hyperplane that minimizes the optimization function is called a {\em QR} Hyperplane and denoted by $\mathcal{H}$
% We call  such a hyperplane {\em QR} Hyperplane and denote it by $\mathcal{H}$. 
As shown in Figure \ref{fig:input_space},for our running example in 2-dimensions, the regression hyperplanes are lines.
The coefficients associated with the {\em QR} Hyperplane are regression parameters ($\beta$).

\stitle{Optimal QR Hyperplane, $\mathcal{H}^*$:} In primal space, there is a {\em QR}  hyperplane for which {\em QR} optimization function is minimum. We call  this hyperplane {\em Optimal QR Hyperplane} and denote it by  $\mathcal{H}^*$.
The {\em Optimal QR Hyperplane} geometrically divides the points into $\tau n$ points on one side and $(1-\tau)n$ points on the other side~\cite{koenker1978regression}.
% On the other hand, the {\em QR hyperplane} ($y=X_i^T\beta$) in the primal space transforms into a point in the dual space. 

% \stitle{{\em QR} Properties:}
The following two important {\em QR} properties are used throughout our paper.
% \vspace{-1mm}
\begin{theorem}\label{thm:convexity} \cite{bloomfield1983least}
The  objective function of QR is continuous and convex.
\end{theorem}
% \vspace{-2mm}
\begin{theorem}\label{thm:opt_k_point} \cite{bloomfield1983least}
% Given a database $\mathcal{D}$ the optimal hyperplane of the quantile regression problem touches at least $d$ points.
% Given a database $\mathcal{D}$ and a value of $\tau \in (0, 1)$ there exists a hyperplane $\mathcal{H}$ with at least $d$ points on it such that the quantile regression optimization function value is the minimum i.e. equal to that of $\mathcal{H}^*$.
The {\em Optimal QR Hyperplane} goes through at least $d$ points.
\end{theorem} 

Note that Theorem~\ref{thm:opt_k_point} immediately suggests the strongly polynomial naive baseline algorithm for the $d$-dimensional {\em QR} problem mentioned in Table~\ref{tab:summary-prior-work}: enumerate all the $\mathcal{O}(n^d)$ hyperplanes that pass through each $d$-sized subset of points of the dataset, compute the objective function in $\mathcal{O} (n)$ time for each hyperplane, and pick the minimum. This simple algorithm runs in $\mathcal{O}(dn^{d+1})$ time.
% \textcolor{red}{Gautam: I described the baseline algorithm right here. Not sure I have teh time complexity right. it is different from what is stated in table 2.}

\vspace{-1mm}
\stitle{Dual space}:
A duality transformation function transforms a given point (hyperplane resp.) in primal space into a hyperplane (point resp.) in the dual space, such that certain properties are maintained in the dual space and vice-versa.
Following along similar lines as Edgeworth~\cite{edgeworth1888xxii}, we define the duality transformation function as follows:
% in primal space into the dual space.
Given a hyperplane $\mathcal{H}$ in $d$ dimensional primal space, the dual of $\mathcal{H}$ (point resp.) is a point   $F(\mathcal{H})  \in \mathbb{R}^d$ (hyperplane $\mathcal{F}(p)$ resp.) such that:

% \vspace{-3mm}
% \capstartfalse
\begin{table}[!ht]
    \centering
    \begin{tabular}{|c|c|}
        \hline
        Primal space & Dual Space \\
        \hline
        $\mathcal{H}: y - a_1 x_1 - a_2 x_2 \ldots - a_{d-1} x_{d-1} - a_d = 0$ & $\mathcal{F}(\mathcal{H}): \{a_1, a_2, \ldots, a_d\} \in \mathbb{R}^d$ \\
        $p: \{p_1, p_2, \dots, p_d\} \in \mathbb{R}^d$ &  $\mathcal{F}(p): z_d + \overset{d-1}{\underset{i=1}{\sum}}p_i z_i - p_d = 0$\\
        \hline
    \end{tabular}
    % \caption{Duality transform between primal and dual space}
    % \label{tbl:primal-dual-functions}
\end{table}
% \vspace{-3mm}

% Given a hyperplane $\mathcal{H}$ in $d$ dimensional primal space, the dual of $\mathcal{H}$ is a point  $F(\mathcal{H}) \in \mathbb{R}^d$ such that:
% \begin{align*}
%    \mathcal{H}&: y - a_1 x_1 - a_2 x_2 \ldots - a_{d-1} x_{d-1} - a_d = 0\\
%    \mathcal{F}(\mathcal{H})&: \{a_1, a_2, \ldots, a_d\} \in \mathbb{R}^d
% \end{align*}

% Given a point $p$ in $d$ dimensional primal space, the dual of the point is a $d$-dimensional hyperplane $\mathcal{F}(p)$ such that:
% \begin{align*}
%    p&: \{p_1, p_2, \dots, p_d\} \in \mathbb{R}^d\\
%    \mathcal{F}(p)&: z_d + \overset{d-1}{\underset{i=1}{\sum}}p_i z_i - p_d = 0
% \end{align*}

Here $z_i$ represents the variables defining the hyperplane in dual space.
We would like to note that the $n$ points in the primal space transform into $n$ corresponding hyperplanes in the dual space.
% \vspace{-2mm}
% \begin{figure}[tp]
%     \centering
%     % \includegraphics[width=0.45\textwidth]{figures/DualSpace.png}
%     \includegraphics[width=0.31\textwidth]{figures/DualSpace2.png}
%     \caption{Dual space of the Sample Database}
%     \label{fig:dual_space}
%     \vspace{-5mm}
% \end{figure}

% \medskip

% \textcolor{red}{Gautam: So, as I have been saying earlier, I think the earlier stuff on geometry, duality, etc, should come very early, even before you start describing connection with LP, earlier state of art algorithms, etc. In any case, we cannot claim credit for this stuff, as it is really all prior work. The later stuff on arrangements, computational geometry, etc, can come later as a separate subsection.}

\stitle{Residual sets}:
Given a {\em QR} hyperplane with parameters $\beta$ in the primal space, the set of points for which value $y_i$ is greater than or equal to  (resp. lesser than) the predicted value $\hat{y_i}=\beta^{T}X_i$ is denoted by $I^+=\{i: y_i \geq \hat{y}_i\}$) (resp. $I^-=\{j: y_j < \hat{y}_j\}$). 
% \abol{I am lost. You did not even use the hyperplane! you are only looking at the y values!!}
% Given a hyperplane, the set of points for which actual value $y$ is lesser than $\hat{y}$ is denoted by $I^-=\{j: y_j < \hat{y}_j\}$). \abol{no need to repeat. You could merge the two sentences.}
Note that the two sets are mutually exclusive, $I^+ \cap I^- = \emptyset$; $I^+ \cup I^- = \mathbb{U}$. 
% \abol{Are you trying to define half-spaces? If so, why do you not use the standard wording provided in Edelsburner's book and also in the stable/fair ranking (and RRR) papers}
% \suraj{We use these $I^+$ and $I^-$, which are points in the open and closed half-space. It would be confusing to a reader if we introduce a half-space here and not use the half-spaces later. I will update the comment based on what you think.}
Geometrically, the entity $I^+$ represents the set of points that lie on or above the regression hyperplane, and $I^-$ represents points that lie below the hyperplane.

\stitle{Complete Skeleton, $S$}: 
% Now we describe how an arrangement $\mathcal{A}$ of $n$ hyperplane can be represented as a graph.
Given a $d$-dimensional space, the intersection of $d$ hyperplanes is a $0$-dimensional point which is called the vertex.
The intersection of $d-1$ hyperplanes creates a $1$-dimensional line segment which is called an edge. 
The endpoints of an edge consist of two vertices which are called neighbors.
These vertices and edges create a connected graph called Complete Skeleton ($S$). 
The details about the {complete skeleton} can be found in \cite{edelsbrunner1987algorithms}. 

\stitle{$k$-set}:
Given a set of $n$ points in $d$-dimensional space, a $k$-set is a subset of $k$ points that can be separated by a hyperplane from the remaining points. Much is known about $k$-sets, their counts, as well as algorithms for their enumeration and construction, especially in lower dimensional space ~\cite{edelsbrunner1987algorithms}.
For our geometric mapping, consider the $n$ points is given by the data points in the primal space.
As the optimal quantile regression hyperplane $\mathcal{H}^*$ must partition the $n$ points such that $\tau n$ points are on one side and the remaining $(1-\tau)n$ lie on the other, the optimal hyperplane is a separating $k$-set hyperplane.
We exploit these observations to design our deterministic strongly polynomial algorithm for $2$ (and higher) dimensions.
% For a detailed review of the geometric concept of Arrangement and its $k$-th level, please refer to Appendix~\ref{subsec:arrangements-ksets}.

\stitle{Arrangements}:
Given $n$ hyperplanes in $d$-dimensional space, the entire space is partitioned into an {\em arrangement} consisting of $O(n^d)$ $d$-dimensional convex {\em cells}. Much is known about the geometric properties of such arrangements, as well as algorithms for their construction~\cite{edelsbrunner1987algorithms}. 
The $n$ points in the primal space transform into $n$ dual hyperplanes, creating a dissection of the dual space into an arrangement~\cite{edelsbrunner1986constructing}. %, which provides us with an alternate view of the problem.
Figure \ref{fig:dual_space} shows the lines in the dual space for the sample points in Figure~\ref{fig:input_space}.

\stitle{$k$-level of an arrangement:} In an arrangement of $n$ hyperplane, the $k$-level of an arrangement is a set of points $p$  such that the number of lines above $p\leq k-1$ and and the number of lines below $p\le n-k$.  
The black bold lines in Figure~\ref{fig:dual_space} shows the $4$-level of arrangement. 

Further details of duality, arrangements, complete skeleton graph, $k$-sets and $k$-level of arrangement can be found in \cite{edelsbrunner1987algorithms}.

%% file: sections/general.tex
\section{Efficiently Updating Neighbors, and the Use of $k$-Sets}\label{sec:general}
%We start our technical contribution by proposing a geometric scheme for solving quantile regression.
%In the following, first we introduce a subroutine for finding the neighbors of an exterior-point efficiently and propose two warm-up algorithms. 
%Then, we describe and analyze our $k$-set based \twodAlg algorithm.

% it is often required to calculate the objective cost of an exterior point. 
% In these approaches, at each exterior point, the objective cost is calculated from scratch by traversing all the input points of dataset which is computationally expensive.
% % In this section, we develop our first technical contribution \un, the key subroutine utilized by our scalable two-dimensional  algorithm.
% % In particular, we first introduce {\em residual sets}, which will later help us formally connect the geometrical concept of $k$-sets with {\em QR}.

As seen in Section \ref{sec:preliminaries}, a vertex in the complete skeleton graph of the arrangement in dual space refers to a QR hyperplane in primal space.
The optimal solution corresponds to one of these vertices in the complete skeleton graph.
Note that each of these vertices (QR hyperplane in primal space) has an optimization function value associated with it. Computing the optimization from scratch at any vertex takes $\mathcal{O}(nd)$ time, and there are $\mathcal{O}({n^d})$ such vertices. In this section we propose two improvements to this basic brute force approach: (a) design an efficient way to incrementally calculate the optimization function at a vertex from precomputed values at a neighboring vertex (along the complete skeleton graph), and (b) trim the complete skeleton graph to only consider the vertices corresponding to $k$-sets.
\subsection{\un}\label{subsec:main-UN}

%The exterior point based approaches to solve {\em QR} often move from one exterior point to a neighboring exterior point such that the objective cost decreases. 
%\textcolor{red}{Gautam: the above sentence is very vague. There is no good definition of ``exterior point based approaches'' thus far. Also, what do we mean by a ``neighboring exterior point''? Neighboring exterior point of the Linear programming convex polytope surface? or neighboring point in the arrangement in dual space?}
%Note that each exterior point has an optimization function value associated with it. 
%While moving from one exterior point to a neighboring point, the objective function is often recalculated  from scratch by traversing all the input points which is expensive and requires $\mathcal{O}(nd)$ time. 
In this subsection, we propose an \un which can efficiently and incrementally calculate the objective function at a vertex of the complete skeleton graph, by leveraging the calculations performed at a neighboring vertex. % by maintaining only a few \textit{aggregate values}.
The incremental computation only takes  $\mathcal{O}(d)$ time at a vertex, as compared to $\mathcal{O}(nd)$ time if performed from scratch. 
%The \un, when provided with an input vertex along with its optimization function value and a neighboring vertex, updates the optimization function value in $\mathcal{O}(d)$ time.
The update operation relies on maintaining a few aggregate values when moving from one vertex to its neighboring vertex.
To understand the motivation behind the aggregate values, we rewrite the optimization function into two summations.

\vspace{-6mm}
\begin{align*}
    % \underset{\beta\in \mathbb{R}^d}{\min}\ \underset{i\in \{i:y_i \geq   X_i\beta\}}{\sum}\tau\ (y_i - X_i\beta)\ +\ \underset{j\in \{j:y_j < X_j\beta\}}{\sum}(1-\tau)\ (X_j\beta - y_j)
    \underset{\beta\in \mathbb{R}^d}{\min}\ \tau\underset{i\in I^+}{\sum}\ (y_i - X_i\beta)\ +\ (1-\tau)\underset{j\in I^-}{\sum}\ (X_j\beta - y_j)
\end{align*}
\vspace{-4mm}

% The points in the primal space transform into hyperplanes in the dual space.
% Each $d$ of these hyperplanes intersect to form points, which we refer to as intersection points.
% The arrangement of the $d$ hyperplanes forms a total of $\binom{n}{d}$ intersection points.
% We refer to two dual intersection points as neighbors if the two points are neighbors in the arrangement, i.e. there are no intersection points on the line that connects the two points.
% Note that each of these intersection points $p$ are hyperplanes $\mathcal{F}^{-1}(p)$ in the primal space.

First, we will show that if we are given certain aggregate values for a vertex, we can compute the optimization function in $\mathcal{O}(d)$ time.
% Consider that the aggregate values $\sum_{i \in I^+} y_i$, $\forall_{1 \leq m \leq d}\ \sum_{i \in I^+} X_i[m]$, $\sum_{j \in I^-} y_j$ and $\forall_{1\leq m \leq d}\ \sum_{j \in I^-} X_j[m]$ were known for a given vertex in the skeleton graph (hyperplane in primal space).
For all $i \in I^+$ and $j \in I^-$, consider the aggregate values $\sum y_i$, $\sum X_i[1]$, $\dots$, $\sum X_i[d]$, $\sum y_j$ and $\sum X_j[1]$, $\dots$, $\sum X_j[d]$ were known for a given vertex in the skeleton graph (hyperplane in primal space).
Given these aggregate values, the optimization function can be computed for the corresponding primal hyperplane in $\mathcal{O}(d)$ time using the following formula.

\vspace{-5mm}
\begin{align}
    \tau\underset{i\in Y^+}{\sum}y_i -\  \tau\underset{1 \leq m \leq d}{\sum}(\ \beta_m\underset{i\in Y^+}{\sum}X_i[m]\ )\ +\ (1-\tau)\underset{1 \leq m \leq d}{\sum}(\ \beta_m\underset{j\in Y^-}{\sum}X_j[m]\ )  - (1-\tau)\underset{j\in Y^-}{\sum}y_j
\end{align}
\vspace{-4mm}

Secondly, we show that, given a vertex and aggregate values corresponding to the vertex, the aggregate values for any neighboring vertex can be computed in $\mathcal{O}(d)$ time.
% $\forall_{1\leq m \leq d}\ \sum_{i \in Y^+} X_i[m]$, $\sum_{i \in Y^+} y_i$ for $I^+$ (the points above the hyperplane) and aggregate values  $\sum_{j \in Y^-} y_j$ and $\forall_{1\leq m \leq d}\ \sum_{j \in Y^-} X_j[m]$ for $I^-$ (the points below the hyperplane can be computed in $\mathcal{O}(d)$ time) for any neighboring point.
An illustration to highlight this result is provided in Section \ref{subsec:illustrate-UN}.
The formal theorem for this $\mathcal{O}(d)$ time update is presented in Theorem \ref{thm:constant_update}.
% The proof is presented in Appendix~\ref{sec:lemmaApp}.

\begin{theorem}\label{thm:constant_update}
Given the aggregate values $\forall_{1\leq m \leq d}\sum_{i \in I^+} X_i[m]$, $\sum_{i \in I^+} y_i$, \\$\sum_{j \in I^-} y_j$, $\forall_{1\leq m \leq d}\ \sum_{j \in I^-} X_j[m]$ for a vertex in the complete skeleton graph, the aggregate values can be updated in $\mathcal{O}(d)$ time when we move to a neighboring vertex.\footnote{All proofs are provided in Appendix \ref{sec:lemmaApp}.}
\end{theorem}

A detailed illustration of \un is presented in Appendix~\ref{subsec:illustrate-UN}.

%% file: sections/2d_soln.tex
% \section{Quantile Regression in 2 Dimension}\label{sec:2d}

% In this section, we start with an brief discussion on approaches that rely solely on the \un oracle or on a neighborhood based exploration of Theorem~\ref{thm:opt_k_point}.
%In the following, we briefly describe two warm-up algorithms that rely solely on the \un and Theorem~\ref{thm:opt_k_point} for exploration.
%Drawing from the strength and drawbacks of these approaches, we describe our algorithm which combines the Theorem~\ref{thm:opt_k_point} and \un oracle to obtain an efficient algorithm for $2$ dimensional QR problem.

\subsection{Improving the Naive Baseline Using \un}
An algorithm that solely relies on \un to find the optimal {\em QR} line can use a neighborhood exploration in the complete skeleton graph $S$.
Initially, a vertex is $S$ is arbitrarily chosen as a start vertex.
In each step of the algorithm, the neighborhood is explored.
% We can start from any of the $\mathcal{O}(n^2)$ vertices in $S$ and explore its  neighborhood.
The vertex which improves the optimization function the most is visited next.
The exploration continues until there are no neighboring vertices with a lower value for the optimization function.
In the worst case, this algorithm may explore the complete skeleton graph, and its worst case time complexity is therefore $\mathcal{O}({dn^d})$. Nevertheless, this algorithm is strongly polynomial under the assumption that $d$ is bounded, and is an improvement over the naive baseline algorithm.
%the vertices in $S$ before it stops.
% does not have a good pruning ability and can explore a large part of the vertices in $S$.
%Thus, its overall complexity is $\mathcal{O}(n^2)$.
A more detailed explanation for this algorithm is presented in Appendix~\ref{subsec:update-neighbor-approach-2d}.

% \begin{wrapfigure}{r}{0.48\textwidth}
%     \centering
%     \vspace{-25mm}
%     \includegraphics[width=0.48\textwidth]{figures/QRREG2D.png}
%     \vspace{-9mm}
%         \caption{The ordered traversal of the k-level of arrangement}
%         \label{fig:qrregseq}
%      \vspace{-6mm}
% \end{wrapfigure}

\subsection{\twodAlg: Leveraging $k$-Sets in the Two Dimensional QR Problem}
%\stitle{\texorpdfstring{$k$}\ -Set Based Algorithm}: 
Recall from Section~\ref{sec:geom-map} that a $k$-set is a subset containing $k$ points that can be separated from the rest of the $(n-k)$ points by a hyperplane.
An interesting observation is that the optimal {\em QR} hyperplane separates the $n$ points into $\tau n$ and $(1-\tau)n$ points~\cite{portnoy1997gaussian}.
While there are many hyperplanes that separate $n$ points into two parts (with $\tau n$ and $(1-\tau)n$ points), we are interested in the specific hyperplane that provides the lowest optimization score.
Since a $k$-set separate the points into $k$ and $n-k$ points, {\bf the optimal QR hyperplane must correspond to a $k$-set}, where $k=\tau n$.
In other words, the optimal QR hyperplane is among the set of possible $k$-set separating hyperplanes, the one with minimum optimization score.

\begin{wrapfigure}{r}{0.48\textwidth}
    \centering
    \vspace{-5mm}
    \includegraphics[width=0.48\textwidth]{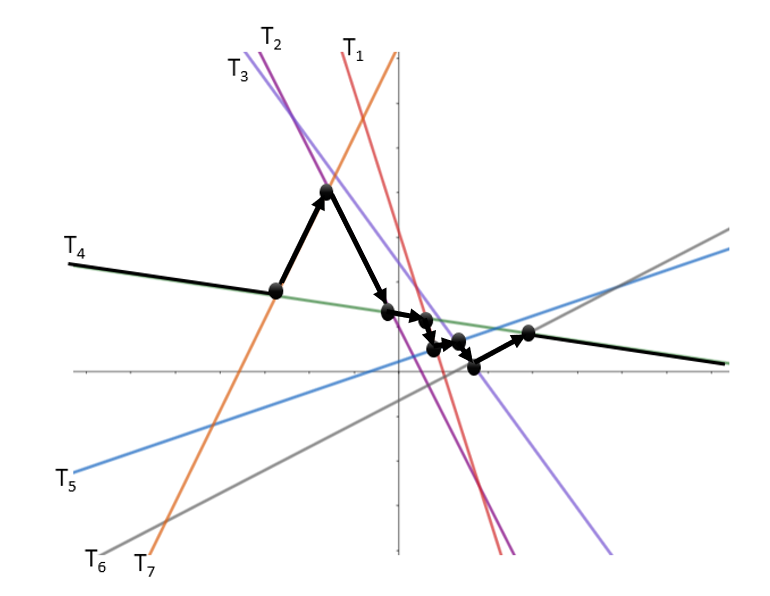}
    \vspace{-9mm}
        \caption{The ordered traversal of the k-level of arrangement}
        \label{fig:qrregseq}
     \vspace{-6mm}
\end{wrapfigure}

For the rest of this subsection, we primarily focus on the 2-dimensional QR problem. 
The enumeration of $k$-sets can also be viewed as a walk along the $k$-level of an arrangement. More specifically, we require the $k$-set enumerating algorithm to provide us with vertices such that any vertex $v_j$ obtained after $v_i$ is its neighbor.
% Note that the neighboring vertices both belong to the $k$-set.
Figure \ref{fig:qrregseq} shows the ordered sequence of $k$-sets for our running example. 
Any $k$-set enumeration algorithm that satisfies this property can be used in our algorithm. 
The best-known deterministic approach is a line-sweep algorithm by Edelsbrunner and Welzl~\cite{edelsbrunner1986constructing} using the dynamic data structure from T. Chan ~\cite{chan2001dynamic}. 
The complexity of this algorithm is $\mathcal{O}(n\log{(m)} + m\log^{1+a}{(n)})$, where $m$ is the number of $k$-set and $a >0$ is an arbitrarily small constant.
Additionally, T. Chan~\cite{chan1999remarks} has developed a randomized incremental algorithm to perform the enumeration in $\mathcal{O}(m + n\log{(n)})$.

The pseudocode of \twodAlg is presented in Appendix~\ref{sec:algorithms-appendix} (Algorithm~\ref{alg:2d-algo}). 
Time complexity of our algorithm with deterministic (randomized resp.) $k$-set enumeration is presented in Lemma~\ref{thm:qrreg2d-proof} (Lemma~\ref{thm:qrreg2d-rand-proof} resp.).
Both the proofs of Lemma \ref{thm:qrreg2d-proof} and Lemma  \ref{thm:qrreg2d-rand-proof} utilize the upper bound of the number of $k$-set \cite{dey1997improved} and $k$-set enumeration techniques \cite{chan1999remarks}. 
% The run time complexity of our algorithm with deterministic $k$-set enumeration is presented in Lemma \ref{thm:qrreg2d-proof}.
% Lemma \ref{thm:qrreg2d-rand-proof} proves the expected run time of our algorithm  when using randomized $k$-sets enumeration.

% Our approach is presented in \twodAlg Algorithm \ref{alg:2d-algo} (in Appendix~\ref{sec:algorithms-appendix}). 
% We initialize the aggregate values $Xp$, $Yp$, $Xm$ and $Ym$ to $0$. 
% Note that $Xp$ and $Yp$ represent the aggregate $\sum X_i$ and $\sum y_i$ for $i \in I^+$.
% Similarly, $Xm$ and $Ym$ represent the aggregate $\sum X_j$ and $\sum y_j$ for $j \in I^-$.
% We compute the aggregate values for the first point in the $k$-set as a special case.
% As we move from one point to its neighboring point in the $k$-set we call \un to update the aggregate values.
% The objective cost of the new point visited can be computed using the aggregate values.
% We return the $k$-set with the minimum objective cost in the walk over the $k$ level of arrangement.
% The run time complexity of our algorithm with deterministic $k$-set enumeration is presented in Lemma \ref{thm:qrreg2d-proof}.
% Lemma \ref{thm:qrreg2d-rand-proof} proves the expected run time of our algorithm  when using randomized $k$-sets enumeration.

\begin{lemma}
\label{thm:qrreg2d-proof}
\twodAlg has a time complexity of 
% Algorithm~\ref{alg:2d-algo} finds the optimal solution for the 2 dimensional {\em QR} problem in 
$\mathcal{O}(n^{\frac{4}{3}}\log^{1+a}{n})$,  where $a >0$ is an arbitrarily small constant\footnote{The proof is provided in Appendix~\ref{sec:lemmaApp}\label{ftnote:App}}.
% time where $a >0$ is an arbitrarily small constant.
\end{lemma}
% \begin{proof}
% For the first $k$-set obtained through the enumeration, the optimization function and aggregate values need to be calculated by a linear scan over the points which takes $\mathcal{O}(n)$ time.
% % The $2$ dimensional algorithm is given in \ref{alg:2d-algo}.
% Updating the optimization function value and aggregate values as we explore neighboring $k$-set takes $\mathcal{O}(1)$ time.
% The points in the $k$-level of the arrangement can be computed using Sweep Line algorithm \cite{edelsbrunner1986constructing, chan2001dynamic, chan1999remarks} in $O(n\log{m} + m\log^{1+a}{n})$ where $m$ is the number of $k$-set and $a >0$ is an arbitrarily small constant.
% The upper bound on total number of $k$-sets is given by Dey~\cite{dey1997improved}, $\mathcal{O}(nk^{1/3})$.
% This brings the overall time taken to $\mathcal{O}(nk^{1/3}\log^{1+a}{n})$.
% As $k$ is a percentage of $n$, the overall time complexity $\mathcal{O}(n^{4/3}\log^{1+a}{n})$.
% Hence, proved.
% \end{proof}

\begin{lemma}
\label{thm:qrreg2d-rand-proof}
\twodAlg with randomized $k$-set enumeration~\cite{chan1999remarks} has an expected time complexity of
% Algorithm~\ref{alg:2d-algo} finds the optimal solution for the 2 dimensional {\em QR} problem in 
$\mathcal{O}(n^{\frac{4}{3}})$\footref{ftnote:App}.
% expected time.
\end{lemma}
Interestingly, as also shown in Tables \ref{tab:summary-prior-work} and \ref{tab:summary}, our $k$-set based algorithm for two dimensions {\em is asymptotically better than} any other existing exterior or interior point approaches in two dimensions ({\em BR} or {IPM}). Extending to higher dimensions is an open problem, as the corresponding problems of counting and enumerating $k$-sets in higher dimensions are generally challenging and open problems in computational geometry. Any breakthroughs are likely to have implications for our $k$-sets based algorithm for QR.

%% file: sections/zpp_approach.tex
\section{\randQR: An Efficient Randomized Algorithm for Quantile Regression}\label{sec:qr-divide-conquer}

In this section, we introduce a more efficient randomized algorithm \randQR for solving the Quantile Regression problem.
The high level overview of our approach and the intuition behind it are described in Section~\ref{subsec:general-zpp-approach}.
We prove its expected running time of $\mathcal{O}(n\log^2{(n)})$ in two dimensions in Section~\ref{subsec:2d-zpp-approach}.
The extension of our approach to three and higher dimensions is presented in Section~\ref{subsec:3d-zpp-approach}.

\subsection{\randQR  Based on a Randomized Divide-and-Conquer Approach}\label{subsec:general-zpp-approach}

As discussed in in Section~\ref{sec:geom-map}, a hyperplane in the primal space transforms to a point in the dual space.
From Theorem~\ref{thm:opt_k_point}, we can deduce that the optimal QR hyperplane is {\em special} as it is located at a vertex in the skeleton graph $S$ of dual space.
A naive search of vertices in $S$ to find the optimal QR hyperplane is computationally expensive as there are $n^d$ vertices.
Our approach relies on restricting the search space to find the optimal QR vertex in $S$ efficiently.

All our restrictions to the search space will be placed on a single variable, called the {\em search variable}.
Initially, an arbitrary variable $A'$ is selected as the {\em search variable}.
That is $A'=A_i$, for an arbitrary value of $i\in[1,d]$.
% \textcolor{red}{Gautam: inconsistent notation. Variables are A, or X?}
% consider $A_i$ to be the search attribute ($A'=A_i$ where $1 \leq i \leq d$). %\abol{is $i$ is one of the values in that range, e.g., $2$, or is it variable? why do you even need to bring $i$ to the picture?} \abol{Why not saying "For the sake of simplicity, let $A'=A_1$."}

% \begin{figure*}[t] 
% \centering
%     % \begin{minipage}[t]{0.40\linewidth}
%         \centering
%         \includegraphics[width=0.40\textwidth]{figures/optimisation_function_one_cut.pdf}
%     \caption{Optimization function of a sample 2D QR problem with the first splitting-plane}
%     \label{fig:optimization-one-cut}
    % \end{minipage}
    % \hfill
    % \begin{minipage}[t]{0.40\linewidth}
    %     \centering
    %     \includegraphics[width=0.98\textwidth]{figures/optimisation_function_two_cut.pdf}
    %     \caption{Optimization function of a sample 2D QR problem after the second iteration}
    %     \label{fig:optimization-two-cut}
    % \end{minipage}
% \end{figure*}

\begin{wrapfigure}{r}{0.40\textwidth}
    \centering
    \vspace{-10mm}
    \includegraphics[width=0.40\textwidth]{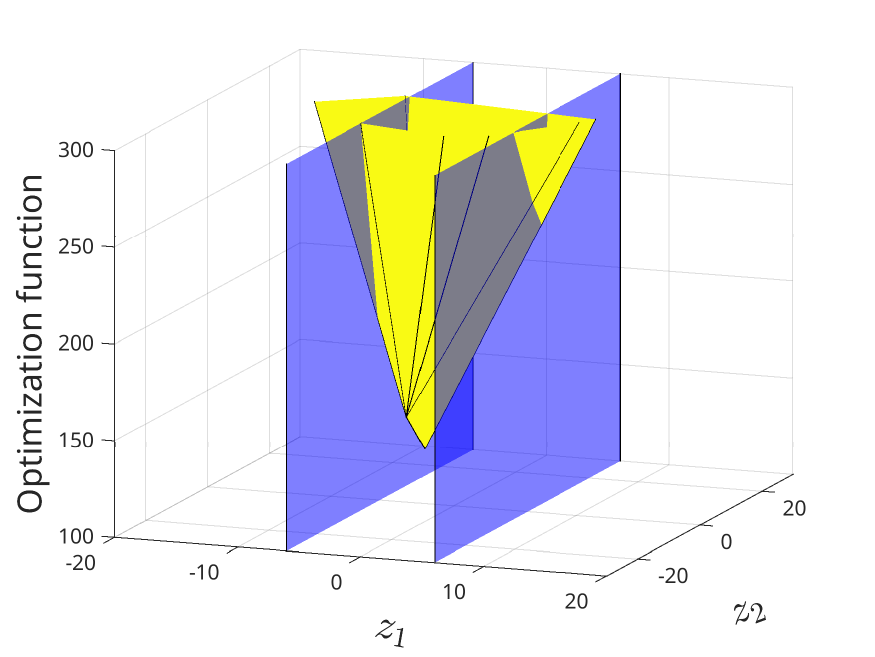}
    \vspace{-6mm}
    \caption{Optimization function of a sample 2D QR problem after the second iteration}
    \label{fig:optimization-two-cut}
     \vspace{-6mm}
\end{wrapfigure}

Our randomized approach breaks the problem into sub-problems based on the divide-and-conquer paradigm. Particularly, starting from $A'\in (-\infty, \infty)$, it keeps restricting the so-called {\em search interval} to smaller ranges denoted as $R=(R_s, R_e)$.
For example, Figure~\ref{fig:optimization-two-cut} shows an example where $A'=A_1$ and $R=(-5,5)$, specified as the range between the two (blue) planes $z_1=-5$ and $z_1=5$\footnote{Recall from Section~\ref{sec:geom-map} that $z_i$ is the $i$-th coordinate in the dual space, corresponding with the variable $A_i$.}.
A vertex $P_j$ would be in the search space if the value of $P_j$ for variable $A'$ lies within the range $R$.
The set of all vertices of $S$ that lie within the search interval are called {\em \belongR}($S_{in}=\{x\in S\: |\: x[A']\in R\}$).
At every iteration, 
the algorithm selects a value $v\in R$ to split the search interval $R$ in three disjoint intervals - $(R_s, v)$, $[v, v]$ and $(v, R_e)$.
The hyperplane that splits $R$ into the three intervals is called the {\em splitting hyperplane}, which is defined as $z_i=v$.

% The interval that restricts the search attribute $A'$ is called the {\em search interval} and is denoted by $R=(R_s, R_e)$.
% A vertex $P_j$ would be in the search space if the value of $P_j$ for attribute $A'$ lies in R.
% Set of all vertices of $S$ that lie in the search interval are termed as {\em \belongR}($\{x\in S\: |\: x[A']\in R\}$).
% As there are no restrictions during the start of our algorithm the search interval for $A'$ is set to $(-\infty, \infty)$.

% Our randomized approach breaks the problem into sub-problems based on the divide-and-conquer paradigm.
% At every iteration, a value $v\in R$ is chosen such that it splits the search interval $R$ in three disjoint intervals - $(R_s, v)$, $[v, v]$ and $(v, R_e)$.
% The hyperplane that splits $R$ into the three intervals is termed as the {\em splitting hyperplane} and has an equation of the form $z_i=v$.

The choice of $v$ during the {\em divide} step gives rise to different nature of algorithms.
For instance, if the interval $R$ were to be divided into three parts based on the mid-value of the range (i.e., $v=(R_s+R_e)/2$), no matter how fast or accurate the conquer step is designed to be, the approach could only lead to a weakly-polynomial algorithm.
Under {general positioning assumption}, the plane $z_i=v$ might contain at most one vertex of $S$.
Hence, to design a strongly polynomial algorithm, ideally the search interval should be divided {\em in a way that \belongR are (almost) split in half}.
% divided such that the  vertices need to be well split into these intervals.
% Hence, to design a efficient strongly polynomial approach a suitable $v$ needs to be chosen such that the \belongR must be split into two ``{\em fairly}'' equal parts - $(R_s, v)$ and $(v, R_e)$.
That is, $|S_{in}\cap(R_s, v)|\simeq |S_{in}\cap(v, R_e)|$.

Instead of selecting $v$ from the continuous range $R$, \randQR uses a vertex from the set of \belongR in order to perform the split. I.e., it identifies a vertex $P\in S_{in}$ for splitting, then uses the value of $P$ on A' as the splitting value (thus $v=P[A']$).
The challenge, however, is that we would like to identify $P$ such that it cuts $S_{in}$ into two equal halves.
Since $|S_{in}|$ is in $\mathcal{O}(n^d)$, enumerating $S_{in}$ for finding $P$ is not feasible.
% if $P_j$ were to be chosen as a vertex \randQR uses the value for $A'$ to decide the value for the cut\abol{this sentence is not complete. I don't understand this and the next sentence}.
% As the set of vertices in the set of \belongR can consist of a large number of vertices of $S$ ($\mathcal{O}(n^d)$ in $d$ dimensions).\abol{not clear what you mean}
% However, 
We leave the problem of efficiently finding the split point using a deterministic algorithm without enumerating $S_{in}$ as an open problem.
% that equally divides the search interval without enumerating a large part of \belongR is a challenging problem, which we leave as an open problem.
% Instead, we propose design a randomized approach for finding the split point.
% Finding a split of almost equal parts for which $R$ is large, without enumerating a large part of \belongR is a challenging problem of its own. We leave that as an open problem which warrants for  research of its own.
Instead, we draw inspiration from the classical {\em Randomized Quick Sort} (RQ-sort) algorithm for identifying the split point.
The randomized pivot selection is a key component of RQ-sort to reduce the quadratic worst-case time complexity of quick sort to linearithmic.
% which has the added advantage that the pivot selection is not a pre-computation step. Furthermore, it is also shown that RQ-sort does not add a lot of overhead to the expected run time of the algorithm compared to the deterministic version of the algorithm.
Applying a similar idea, at a high level, \randQR 
aims to select the split vertex $P$ unformly at random from $S_{in}$. 
The interval $R$ is partitioned into three disjoint intervals, and the interval which contains the optimal vertex is identified.
Consequently, $R$ is updated to the new, smaller interval.

Based on our randomized divide-and-conquer approach, there are two key {\em questions}.
These questions are addressed in the subsequent sections where we discuss them in detail.

\begin{itemize}
    \item ({\em Divide step}) 
    A major difference between RQ-sort and our problem is that, unlike RQ-sort which has random-access to the elements of an array, in our problem {\em $S_{in}$ is not materialized}. As a result, even the exact size of $S_{in}$ is not known apriori. Hence, it is not possible to (a) directly generate a random index in range $|S_{in}|$ and (b) have a random access to a vertex for the generated index. Therefore, a key question is how to efficiently sample (uniformly at random) a vertex from the set of \belongR?
    \item ({\em Conquer step}) How to find out which among the 3 intervals contain the optimum vertex?
\end{itemize}

% In each iteration, a vertex $P_j$ in $S$ that satisfies the restriction on $A'$ is chosen uniformly at random.
% Let the value of $P_j$ for attribute $A'$ be $v$.
% Search interval $R$ is split into three disjoint parts based on $v$ - $(R_s, v)$, $[v, v]$ and $(v, R_e)$.
% Our approach relies on a Oracle which we call \SplitOrc that determines the interval in which the optimal vertex can be found.
% Note that if the \SplitOrc returns the interval $[v,v]$, the vertex $P_j$ is the optimal vertex (hyperplane) for the QR problem.
% Based on the interval determined by the \SplitOrc we update the search interval $R$.
% The details of the \SplitOrc and the procedure to choose a vertex uniformly at random are both discussed in later sections.

% When a small number of vertices are left that satisfy the restrictions on $A'$, optimization value for each of these vertices is computed to find the optimal QR hyperplane.
We design two functions \SplitFn \ and \ConsSamplFn\ to address these two questions.
The role of the \ConsSampl function is to sample a vertex uniformly at random from the set of \belongR.
\SplitOrc finds out the interval among the three intervals where the optimum vertex lies.
The details of both these functions will be discussed in later-part of this section.
The pseudo-code for our approach is presented in Algorithm~\ref{alg:generic-zpp-algo} in Appendix~\ref{sec:algorithms-appendix}.

\subsection{\randQR in two dimensions}\label{subsec:2d-zpp-approach}

\randQR relies on two key components - \SplitFn\  and \ConsSamplFn.
We describe our approaches for both these problems for two dimensions below.

\vspace{2mm}
\noindent{\em First attempt, a Rejection Sampling approach}:
In 2D, a vertex is the intersection of a pair of dual lines. 
Also, following the general positioning assumption, each pair of dual lines intersect exactly once.
As a result, in order to draw a uniform random sample from $S$, one can sample one of the $n \choose 2$ dual lines and compute their intersection in constant-time to find the drawn vertex. 
Therefore, it is straightforward to design a rejection sampling to draw a sample from the set of \belongR $S_{in}$: (1) sample a pair of dual lines, uniformly at random; (2) compute their intersection point $P$; (3) accept $P$ if $P[A']\in R$, otherwise reject it and try again.
This approach, however, is not efficient since the probability of accepting a sample when $|S_{in}|=\mathcal{O}(1)$ is as low as $\frac{1}{n^2}$. As a result, the expected cost of generating one sample for such regions is $\mathcal{O}(n^2)$.

% Initially, there are no restrictions on the search interval, $R$ is set to $A'\in (-\infty,\infty)$, and $S_{in}=S$.
% Hence, any vertex uniformly sampled from $S$ would also be a uniform sampled from $S_{in}$.
% Uniformly sampling any two of the $n$ dual lines produces a vertex among the $\binom{n}{2}$ vertices in $S$.
% The challenge arises in the later steps when $R$ has been updated.
% Once updated, the same strategy would have to account for rejecting a sampled vertex which does not satisfy $R$.
% This {\em rejection sampling} strategy follows a geometric distribution with exponentially diminishing returns.

\stitle{\ConsSampl}: 
Given the inefficiency of the rejection sampling approach, in the following we propose an efficient approach that only generate samples (uniformly at random) that are already in $S_{in}$ and, hence, no rejection is needed.
% Instead of a rejection sampling approach, our strategy is to do a weighted sampling on the lines.
Our strategy is based on weighted sampling of the dual lines, where the weight of each dual line is the number intersections it has within the search interval $R$.
% The weights of each line represents the number of intersections that the line was involved-in within the interval $R$.
The weights are then used to sample one line.
% The probability of sampling a line $\ell_i$ is then $w_i/(\sum_{k=1}^n w_k)$.
Once a line is sampled, all the vertices of $S_{in}$ that involve $\ell_i$ will be enumerated and one of those vertices are returned uniformly at random.
% One vertex can be uniformly sampled among the enumerated vertices.

\begin{lemma}\label{lem3}
\ConsSampl is an unbiased sampler for $S_{in}$\footnote{The proof is provided in Appendix~\ref{sec:lemmaApp}}.
\end{lemma}

% \begin{proof}
%     Let $S_{in}^{(i)}$ be the number of intersections in $S_{in}$ that involve $\ell_i$.
%     The probability of selecting each line $\ell_i$ is $$Pr(\ell_i)= \frac{|S_{in}^{(i)}|}{2|S_{in}|}$$
%     After selecting a line $\ell_i$, the probability that a specific intersection $P_j$ involving it is $\frac{1}{|S_{in}^{(i)}|}$.
%     Now, for an intersection $P_j\in S_{in}$, let $\ell_i$ and $\ell_k$ be the dual lines that involve it.
%     The probability of sampling $P_j$ is 
%     \begin{align*}
%         Pr(P_j)&= \frac{|S_{in}^{(i)}|}{2|S_{in}|}\times \frac{1}{|S_{in}^{(i)}|} + \frac{|S_{in}^{(k)}|}{2|S_{in}|}\times \frac{1}{|S_{in}^{(k)}|} \\
%         &= \frac{1}{|S_{in}|}
%     \end{align*}
% \end{proof}

% \begin{wrapfigure}{r}{0.22\textwidth}
%     \centering
%     \vspace{-15mm}
%     \includegraphics[width=0.22\textwidth]{figures/Inversions2D.pdf}
%     \vspace{-6mm}
%     \caption{Depicting the start and end order in a sample 2D QR problem}
%     \label{fig:inversions-2D}
%      \vspace{-6mm}
% \end{wrapfigure}

\begin{figure*}[t] 
\centering
    \begin{minipage}[t]{0.23\linewidth}
        \centering
        \includegraphics[width=0.98\textwidth]{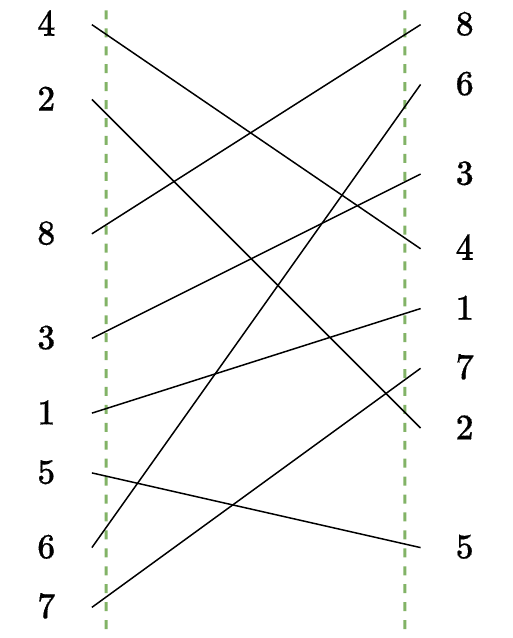}
    \caption{Depicting the start and end order in a sample 2D QR problem}
    \label{fig:inversions-2D}
    \end{minipage}
    \hfill
    \begin{minipage}[t]{0.43\linewidth}
        \centering
        \includegraphics[width=0.98\textwidth]{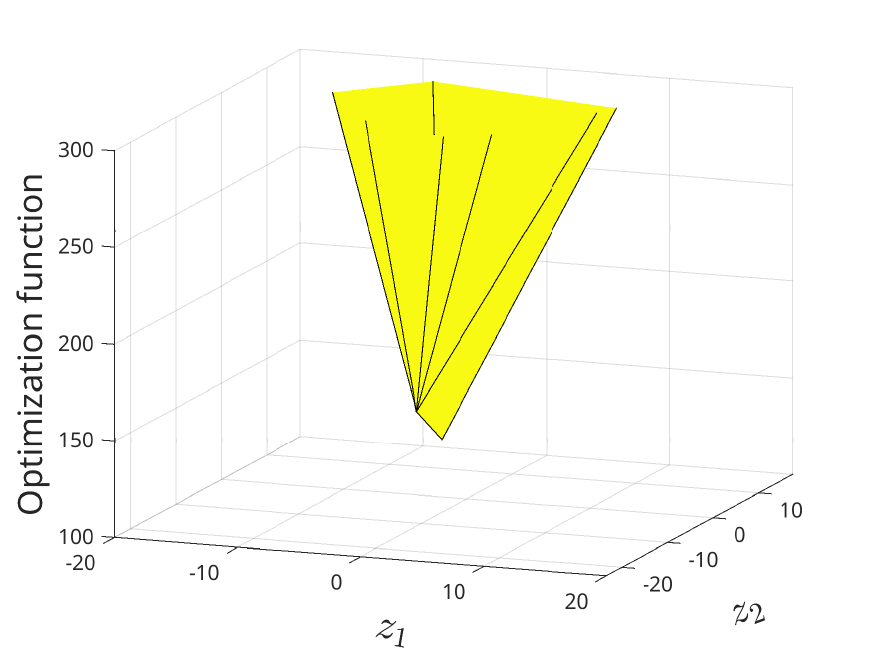}
        \caption{Optimization function of a sample 2D QR problem
        % \abol{change the axis-labels to z1, z2}
        }
        \label{fig:optimization-function}
    \end{minipage}
    \hfill
    \begin{minipage}[t]{0.27\linewidth}
        \centering
        \includegraphics[width=0.98\textwidth]{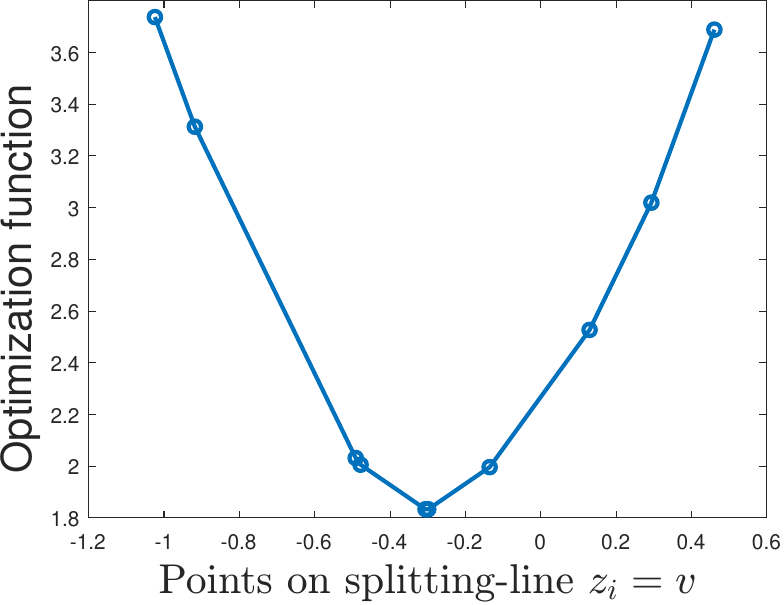}
        \caption{The optimization function over the splitting line in 2D
        % \abol{what are the x,y axis labels?}
        }
        \label{fig:splitting-line-optimiztion-function}
    \end{minipage}
\end{figure*}

We still need to find the weight of each dual line $\ell_i$, that is, the number of vertices in $S_{in}$ involving $\ell_i$.
% Thus, the problem of sampling a vertex $S_{in}$ depends on finding the number of intersections in $R$ that involve each line $\ell_i$.
% Our sampling strategy relies on obtaining the vertex-count that each of the $n$ dual lines are involved in.
% Only the vertices that satisfy the restrictions $R$ on attribute $A'$ are considered for the count.
% The counts are then used as weights in the sampling strategy to obtain the lines that intersect at a vertex.
% Consider the search interval $R=(R_s, R_e)$ in one of the iterations.
% A simple observation is that every line passes through the interval $R$ and there are no vertical lines.
% A commonly used approach to enumerate the intersections within an interval is the {\em sweep line }~\cite{shamos1976geometric} algorithm.
% % The vertices that lie within $R$ are formed by the intersections that each of these lines are involved-in within the interval $R$.
% Naively computing the intersections using the {\em sweep line } approach is computationally expensive as the number of intersections may still be in the order of large ($\mathcal{O}(n^2)$).
% We would like to point out that the {\em sweep line} approach not only counts the vertices within $R$, it also enumerates them.
% As the sampling approach only requires computing the counts of intersections within the interval $R$, we propose a different approach. 
Consider the vertical line $z_1=R_s$ and the order in which the $n$ lines intersect with it.
The vertical line that is placed at the start (end, resp.) of the interval $R$ is termed as {\em start border} ({\em end border}, resp.).
Let this order (top to bottom) in which the lines intersect with the {\em start border} be named as {\em start order}.
Similarly, {\em end order} is the order of intersection of the {\em end border}  with the $n$ lines.
In Figure~\ref{fig:inversions-2D}, the left and right vertical boundary lines in show the start and end borders in that example, while the ordering of dual line intersections with them are the start and end orders.
Note that the order of the lines at $\infty$ ($-\infty$ resp.) can be obtained by sorting the slopes of the lines in ascending (descending resp.) order.
% \abol{this needs to get improved. Will get back to it after the corresponding figure is added}

% Based on these two orderings, we make an observation.
Consider the two orderings. Suppose the intersection of the line $\ell_i$ is blow line $\ell_j$ on the left line, i.e., $\ell_i<_{l}\ell_j$, while this ordering is reverse in the right line, i.e., $\ell_j<_{r}\ell_i$.
In this situation the two lines $\ell_i$ and $\ell_j$ must intersect somewhere in the search range $R$.
Also, the pairs of lines that their ordering do not change in the two lists must not intersect in the search range $R$.
% As a result, the size of \belongR $|S_{in}|$ is equal to the number ``{\em inversions}'' between the two lists.
As a result, the size of \belongR $|S_{in}|$ is equal to the number ``{\em inversions}'' in a permutation.
% \textcolor{red}{Gautam: is it known as inversions in a permutation? This is what I have mentioned in the introduction.}

% If a line $P_i$ was placed above $P_j$ in the start order and is placed below $P_j$ in the end order, there has been an intersection between the two within $R$. We call such exchanges as {\em inversions}. For every line $P_i$, the goal is to find the number of such {\em inversions} that have occurred between the start and end orders.

% \begin{definition}
%     Inversions in permutation: Given a permutation of elements, a pair of elements ($e_i$,$e_j$) is called an inversion if $i > j$ and $e_i<e_j$.
% \end{definition}

% Counting the number of inversions between two lists is a well-known textbook example for which the divide and conquer algorithm achieves the time complexity of $\mathcal{O}(n\log n)$~\cite{kleinberg2006algorithm}.
Counting the number of inversions in a permutation is a well-known textbook example for which the divide and conquer algorithm achieves the time complexity of $\mathcal{O}(n\log n)$~\cite{kleinberg2006algorithm}.

% The concept of {\em inversion} in permutations has been well studied~\cite{chan2010counting}.
% The start and end orders from our problem form the two permutations for the counting inversion problem. 
% Chan et. al~\cite{chan2010counting} perform ``vertically partitioning'' of a trie to propose a $\mathcal{O}(n\sqrt{\log{(n)}})$ method to count inversions.

% Inversions can be counted using a merge procedure, which takes a total of $\mathcal{O}(n\log{(n)})$ time.

% Our approach exploits the order of the lines at the two ends of the search interval to compute the counts directly.
% Consider the vertical line $y=R_s$ ($y=R_e$ respectively) and the order in which the $n$ dual lines intersect with it.
% The order is computed by sorting these intersections.
% Note that the order of the lines at $-\infty$ ($\infty$ resp.) can be obtained by sorting the slopes of the lines in ascending (descending resp.) order.

% Let the order of the intersections on the start (end resp.) be denoted as $E_s$ ($E_e$ resp.).
% Two lines have intersected in the interval $R$ if they have exchanged position in the two orders - $E_s$ and $E_e$.
% Such an exchange is called an {\em inversion}.

% TODO
% \suraj{Add illustration here.}

% \begin{theorem}\label{thm:inversions-permutation-run-time}
% Given two permutations of $n$ numbers, $P_1$ and $P_2$, the total time needed to compute the number of exchanges to transform $P_1$ into $P_2$ is $\mathcal{O}(n\log{(n)}))$
% \end{theorem}

\stitle{The \SplitOrc}: 
Given the splitting line $z_1=v$, the task of \SplitOrc is to determine if optimal solution lies on the line
, and if not, identify if it is to the left  or to the right of the splitting line (i.e. in $(R_s, v)$ or $(v, R_e)$).
% If the optimal solution does not lie on it, \SplitOrc determines if the optimal solution lies to the left  or to the right of the splitting line (i.e. in $(R_s, v)$ or $(v, R_e)$).

\begin{wrapfigure}{r}{0.40\textwidth}
    \centering
    \vspace{-9mm}
    \includegraphics[width=0.40\textwidth]{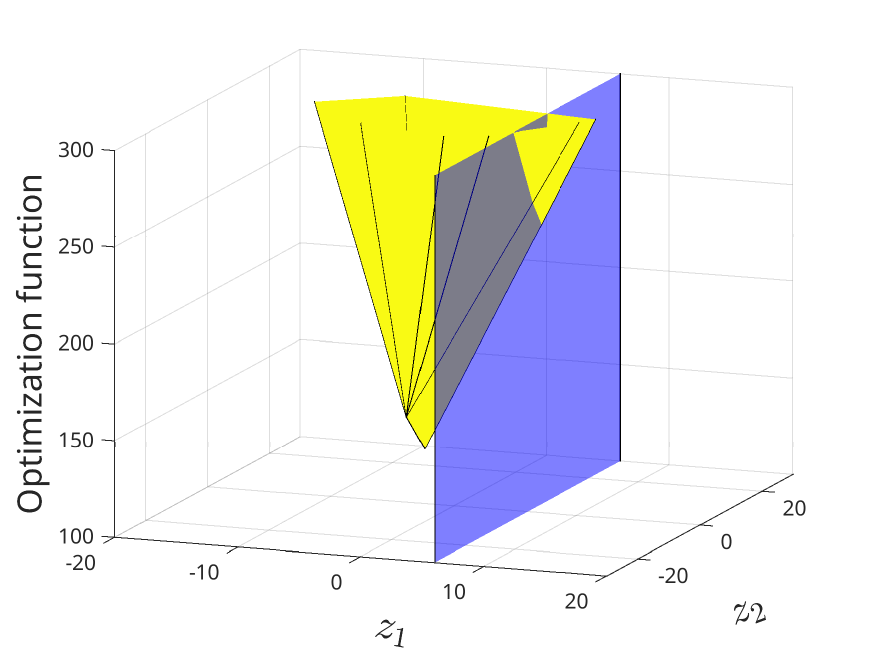}
    \vspace{-6mm}
    \caption{Optimization function of a sample 2D QR problem with the first splitting-plane}
    \label{fig:optimization-one-cut}
     \vspace{-4mm}
\end{wrapfigure}

% \begin{figure*}[t] 
% \centering
%     % \begin{minipage}[t]{0.40\linewidth}
%         \centering
%         \includegraphics[width=0.40\textwidth]{figures/optimisation_function_one_cut.pdf}
%     \caption{Optimization function of a sample 2D QR problem with the first splitting-plane}
%     \label{fig:optimization-one-cut}
%     % \end{minipage}
%     % \hfill
%     % \begin{minipage}[t]{0.40\linewidth}
%     %     \centering
%     %     \includegraphics[width=0.98\textwidth]{figures/optimisation_function_two_cut.pdf}
%     %     \caption{Optimization function of a sample 2D QR problem after the second iteration}
%     %     \label{fig:optimization-two-cut}
%     % \end{minipage}
% \end{figure*}

In order to better describe our solution, consider the dual space being augmented by a third dimension, where the third dimension represents the QR optimization function value (e.g., Figure~\ref{fig:optimization-function}).
The curve defined in the new space is a {\em convex} shape with piece-wise plane faces.
The optimization value for the edges that connect the vertices in $S$ form the boundaries for these plane faces (e.g., Figure~\ref{fig:optimization-one-cut}).
At the vertex where the optimal solution is located, the {\em convex} curve (e.g., Figure~\ref{fig:splitting-line-optimiztion-function}) reaches the least possible value.
% \textcolor{red}{Gautam: general comment ... try and put the figures very close to where they are being described. Wrap text around them if necessary. Another general comment .... you have too many little paragraphs. Try and combine some of them into bigger paragraphs}

% \suraj{Add example here.}\abol{I guess it is clear now from the figure}

The value $v$ for variable $A_1$ is used to partition the augmented dual plane into three parts on $z_1$ axis, the half-plane $(R_s, v)$, the line - $[v, v]$ and the half-plane $(v, R_e)$.
The first question to answer is {\em if the optimal solution lies on the splitting-line?}

Our design of the \SplitOrc for two dimensions relies on the convexity property of the optimization function (Theorem~\ref{thm:convexity}).
From the theorem, we can deduce that the optimization function is convex within the splitting-plane.
Based on convexity, we deploy a divide-and-conquer approach to reach the minima in the plane.
The details of our approach are described below.

The vertical splitting plane intersects the dual lines at $n$ points in the plane. Recall that the ordering of these intersections is called the split order.
% Let the order in which the lines intersection with the plane be called as {\em split order}.
As the curve is convex in the plane,  the point where the minima is reached must correspond to one of these $n$ points.
At the minima-point the optimization function score is least within the plane.

At a given point in the split order, the slope of the optimization function can be computed using the optimization scores of the neighboring points.
Based on this score, the {\em minima-point} has a unique property that both the neighboring points in the split order must have larger optimization values (see Figure~\ref{fig:splitting-line-optimiztion-function}).
Our approach is based on this property that the optimization scores in either direction of the {\em minima-point} are monotone and increasing.
Note that the slopes on either side keep growing in magnitude but have opposite signs.

Our divide-and-conquer approach uses a binary search with two pointers on either end.
The sign of the mid-vertex's slope is used as a marker to update the two ends.
The binary search concludes when we have reached a point where the slope on either side have different signs.
The neighborhood of the minima-point can be used to determine if it is the global minima.
If all points in the neighborhood have a higher optimization value than the minima-point, then it is a global minima.
If not, at least one point in the neighborhood must have a lower optimization value.
% As the optimization function is convex, either the minima-point in the split-plane is a global minima, or the optimization values on one of the sides (left or right) will be greater in any direction.
We use this neighborhood property to determine the optimum interval which contains the optimum point.

Consider the line (lines resp.) that has (have  resp.) intersected with the splitting-line at the minima-point.
In order to obtain the neighborhood of these points, we compute the intersection of these lines with the remaining $n-1$ dual lines.
The intersections (vertices) that are closest to the splitting-line on either side are considered as its neighborhood.
If the optimization value of all of these neighborhood points is greater than the minima-point in the plane then minima-point is a global minima.
If the current point is not a global minima, then the optimization scores on one side must be strictly greater in any direction.
That direction along with the plane $[v, v]$ are discarded.

The pseudocodes of \SplitOrc, \texttt{BinarySearch2D}, and \texttt{ComputeIntervals} are provided in Appendix~\ref{sec:algorithms-appendix}, in Algorithms~\ref{alg:split-pointers-2d}, \ref{alg:binary-search-2d}, and \ref{alg:compute-interval-2d}, respectively.

\begin{theorem}\label{thm:expected-runtime-2d-zpp}
The expected time complexity of \randQR in two dimensions is $\mathcal{O}(n\log^2{(n)})$\footnote{The proof is provided in Appendix~\ref{sec:lemmaApp}}.
\end{theorem}

\subsection{Beyond two dimensions}\label{subsec:3d-zpp-approach}

% \begin{theorem}\label{thm:expected-runtime-3d-divide-conquer}
% The expected run-time of our divide and conquer algorithm in three dimensions is $\mathcal{O}(n^2\log^2{(n)})$
% \end{theorem}

In this section, we extend \SplitOrc and \ConsSampl to three and higher dimensions.

% \textcolor{red}{Gautam: this entire subsection of beyond two dimensions may be moved to the appendix. You could simply have a couple of lines at the end of the previous subsection where you give some very high level idea of what needs to be done for >= 2 dimensions,}

\stitle{\SplitOrc}:
Given a hyperplane in $d$ dimensions ($z_i=v$), the key question that \SplitOrc has to answer is: {\em does the (global) optimal solution lie on the splitting-hyperplane?}
If not, does it lie to left of the splitting-plane or the right?

To answer this question, our approach uses a property that captures the relation between hyperplanes in $d$ dimension and  $d-1$ dimensional sub-spaces.
A hyperplane in $d$ dimensions is a sub-space with $d-1$ dimensions.
If $d$ was 3, then the \SplitOrc would want to know the point in the plane where the optimization function is the smallest.
But this is exactly the QR problem in two dimensions.
Similar observations apply for higher dimensions as well.
Hence, as a first step, our solution to the \SplitOrc\ in $d$ dimensions can reuse the \randQR algorithm in $d-1$ dimensions to find the optimal solution within the plane.
% Hence, the first step in the \SplitOrc in $d$ dimensions is to call the \randQR algorithm in $d-1$ dimensions.

Once a optimal solution within the $d$ dimensional splitting-plane is computed, the next task is to check if it is the global optima.
The approach is similar to the two dimensional \SplitOrc.
The neighborhood of the {\em minima-vertex} determines the interval that will be used for the next iteration of \randQR.

Two vertices are neighbors if they share an edge in the skeleton graph $S$.
Note that every vertex is formed by intersection of $d$ hyperplanes.
As edges in $S$ are segments of lines in the $d$ dimensional space, two neighboring vertices must have $d-1$ hyperplanes in common for them to be incident on an edge.
Therefore, to compute the neighborhood of the minima-vertex, the $\binom{d}{d-1}=d$ lines that pass through the minima-vertex are computed.
Consider a line $l_i$ for this explanation.
Each intersection between $l_i$ and the remaining $n-d$ planes gives us a vertex.
Among these vertices, the vertices which are closest to the splitting-plane on the left and right gives us the two neighboring vertices for the line $l_i$.
Similarly, the neighbors along each of the $d$ lines are computed.
% those vertices which are closest to the splitting-plane $z_i=v$.
There are a total of $2d$ neighboring vertices.

For each of these vertices, optimization value is computed in linear ($\mathcal{O}(nd)$) time.
If the optimization value of the minima-vertex is lower than all these $2d$ vertices, then it is a global minima.
If there is a direction which has a lower optimization score, the hyperplane $[v, v]$ and the other interval are discarded.
The pseudocode of \SplitFn\ for three and higher dimensions is presented in Algorithm~\ref{alg:split-pointers-multi-dimensional} in Appendix~\ref{sec:algorithms-appendix}.

\stitle{\ConsSampl}:
The problem of sampling a vertex from the set of \belongR in higher dimensions is a much harder problem due to the exponential growth of the number of vertices with dimensions ($\mathcal{O}(n^d)$).
% Our goal is to uniformly sample vertices from the set of \belongR.
% The sampling approach we design is similar to the one from two dimensions.
% To obtain a vertex $d$ hyperplanes need to be sampled.
% The intersection point is sampled in a $d$ step process.
% In each step, one among the remaining hyperplanes are sampled based on weights.
Our objective is to achieve a uniform sampling of vertices from the set of \belongR.
The sampling technique we have devised resembles the approach designed for two dimensions.
To obtain a vertex, we must sample a total of $d$ hyperplanes.
The intersection point is determined through a step-by-step process involving $d$ steps.
During each step, one of the remaining hyperplanes is sampled based on its respective weight.
Therefore, in each step, it is necessary to calculate the weights required for the sampling process.

\noindent{\em First step}:
% In the first step, the weight of a hyperplane is the number of intersections that a hyperplane generates within $R$.
In the first step, the weight assigned to a hyperplane corresponds to the count of intersections it generates within the range $R$.
To provide a clearer understanding of the process, consider a three-dimensional example.
When examining the plane $P_1$, the remaining planes are represented as lines (typically referred to as $d-2$ dimensional objects).
% When we look at the plane $P_1$, each of the other planes show up as lines (in general $d-2$ dimensional objects).
A simple observation is that any intersection that involves a plane must occur within the range $R$ on that plane.
For example, say planes $P_1$, $P_2$ and $P_3$ intersected to form a vertex within the search interval $R$.
This intersection vertex must lie on all three planes $P_1$, $P_2$ and $P_3$.
Counting the number of Intersections on a plane has already been studied in the two dimensional \ConsSampl approach.
% Hence, in order to count the number of intersections that a plane has within $R$ in three dimensions, the two dimensional \ConsSampl is invoked with the right parameters.
Therefore, to determine the count of intersections a plane has within the range $R$ in three dimensions, the two-dimensional \ConsSampl method is utilized with the appropriate parameters.

% Once a plane's intersections are accounted for we can discard the plane and continue the process.

In general $d$ dimensional space, the $d$ dimensional hyperplanes are $d-1$ dimensional sub-spaces.
% As seen from the three dimensional example, if a hyperplane $P_i$ is has an intersection within $R$ it must lie somewhere on $P_i$.
As observed in the example of three-dimensional space, if a hyperplane $P_i$ has an intersection within the range $R$, it must be located somewhere on $P_i$.
Hence, we can invoke the \ConsSamplFn\ in $d-1$ dimensional space.
Whenever an intersection occurs, the counts of all the hyperplanes that have been a part of the intersection are incremented by one.

Our approach for \ConsSampl in $d$ dimensions goes over each hyperplane and accounts for the intersections that happens on each hyperplane.
% Once all the intersections for a hyperplane $P_i$ are accounted for, $P_i$ is removed from the set of planes and is not consider for further intersections.
Once all the intersections for a particular hyperplane $P_i$ have been considered, $P_i$ is eliminated from the set of planes and is no longer taken into account for subsequent intersections.
At the end of this process, all the hyperplanes are updated with their respective intersections counts.
Based on these counts the first hyperplane is sampled.

\noindent{\em Second and later steps}: 
The process of sampling the next hyperplane follows a similar procedure.
In step $i$, $i-1$ hyperplanes have already been sampled.
The intersection of the sampled hyperplanes forms a $d-i$ dimensional sub-space.
% Sampling the next hyperplane follows a similar process.
There are a total of $n-d+1$ hyperplanes left to sample from.
\ConsSamplFn\ from the $d-i$ dimensions is used in a similar manner as the first step.

The process of sampling these hyperplanes continues until all the $d$ hyperplanes are sampled.
This pseudo code is presented in Algorithm~\ref{alg:sampling-points-multi-dimensional} in Appendix~\ref{sec:algorithms-appendix}.

\begin{theorem}\label{thm:expected-runtime-3d-zpp}
The expected time complexity of our \randQR algorithm in higher dimensions ($d \geq 3$) is $\mathcal{O}(d\, n^{d-1}\log^2{(n)})$\footnote{The proof is provided in Appendix~\ref{sec:lemmaApp}}.
\end{theorem}

% \textcolor{red}{Gautam: mention where the proof may be found.}

% \suraj{Add a theorem for unbiased sampling?}

% \suraj{ToDo: Update proof}

% \begin{corollary}\label{thm:expected-runtime-3d-zpp-weakly-polynomial}
% The expected run-time of our weakly polynomial \randQR approach in three dimensions is $\mathcal{O}(n\log^2{(n)}L)$ where $L$ is the number of bits used to represent the input data.
% \end{corollary}

%% file: sections/appendix.tex
\section{Quantile Regression: Challenges, and State of Art}\label{subsec:qt-challenges-state-of-art}

% \stitle{Linear Regression $-$ Ordinary Least Squares vs Quantile:}
Linear Regression is a seminal approach in statistics and machine learning, used for building linear predictive models between a response (i.e., dependent) variable and one or more predictor (i.e., independent) variables.
Linear regression is over two centuries old (dating back to Legendre and Gauss) and is often considered as the forerunner of modern machine learning~\cite{portnoy1997gaussian}.
% It is essential to have  a scalable and robust linear regression technique on resource constraint.
We revisit two classical linear regression techniques: {\em Ordinary Least Square Regression} (OLS) and {\em Quantile Regression} (QR). Both build linear models but with different objective functions: in OLS, the objective is to minimize the mean squared error (i.e., $\ell_2$ norm) between the dependent variable and that predicted by the model; whereas in {\em QR}, the objective is to minimize the mean absolute error (i.e., {\em $\ell_1$ norm}).

\stitle{Why Quantile Regression?}
Historically, OLS has been much more widely used than {\em QR}. This is primarily because (as we will discuss shortly) {\em QR} is plagued by unacceptably high computational resource needs, whereas there exist extremely resource-efficient and scalable algorithms for building OLS models.

Nevertheless, despite its computational advantages, OLS has a major limitation:
%It assumes that prediction errors are normally distributed.
% Modern datasets are extremely complex\abol{what do you mean by complex? how is it relevant} and can have an unknown skewed distribution. 
{\em OLS is not a robust model}. It can be easily skewed in the presence of outliers in the data, 
which makes OLS a poor model of choice in several emerging applications where the robustness of the predictive model is critically needed.
On the other hand, {\em  QR is a robust model}. In simple terms, the comparison of OLS and {\em QR} is analogous to the ``mean versus median'' comparison mentioned above. 
%Just like the many applications in which mean is not a robust-enough measurement and hence users are interested in finding the median, 
%OLS is also not reliable in many applications and robust prediction based on Quantile Regression is desirable.
That is why {\em QR} is considered for applications where model robustness in the presence of outliers and skew in the data is critically needed, such as healthcare~\cite{olsen2017use}, ecology~\cite{cade2003gentle}, and others \cite{davino2013quantile}.
Beyond robustness, another attraction of quantile regression is that it is advantageous when conditional quantile functions are of interest, which has application in uncertainty quantification and conformal prediction in AI systems as well as approximate query processing in databases ~\cite{thirumuruganathan2022prediction, savva2020ml}.
% \textcolor{red}{Gautam: give some citations to the above applications}.
In fact, if the scalability and computational efficiency of {\em QR} could be significantly improved, its applicability would increase dramatically across an even more broad range of applications \cite{cade2003gentle}.

\stitle{Computational Challenges of Quantile Regression:}  
% Curious reader might wonder If Quantile Regression is so much robust than OLS, why more people are still using OLS?
%Quantile Regression struggles to face big data when the aspect of speed and memory usage comes.
%On the other hand, OLS is fast and highly scalable, which is the reason behind its popularity compared to Quantile Regression.
% On contrary, Quantile Regression struggles to face big data when the aspect of speed and memory usage comes.
%Unfortunately, Quantile Regression is a much more challenging problem compared to OLS.
%Let us explain with an example.
The computational challenges of {\em QR} are best highlighted by the following observations.
Consider a database with $n$ points and $d$ attributes, of which one attribute is the response (i.e., independent) variable, and the rest are predictor (i.e., dependent) variables.
Typically $n >> d$ in most data fitting problems. For OLS, since the objective function to be minimized (least square) is quadratic and differentiable, this yields a highly scalable training algorithm with time complexity of $O(n d^3)$ time, and space complexity of $O(d^2)$~\cite{chernick2002elements}. Note that the time complexity is effectively linear in $n$ for bounded dimensions.
%is to minimize the sum of the least squared error (i.e., $\ell_2$ norm) by solving a $n$ linear system of equations with $d$ variables.
In contrast, the objective function in quantile regression is defined by the {\em $\ell_1$ norm, i.e., sum of the absolute error}, which has ``piecewise linear'' characteristics.

Hence, its optimization is much more challenging.
% The equivalent of OLS in Quantile regression is median or $ell_1$ regression where the sum absolute error is minimized.
% Due to the absolute operation in objective function, linear programming cannot be applied directly.
%Optimizing for $\ell_1$ norm\abol{make sure to use $\ell_1$ across the paper, not $l_1$} is known to be challenging.
The state-of-the-art approach for {\em QR} \cite{portnoy1997gaussian} applies a reparametrization technique to convert the problem into a huge linear programming formulation involving $n$ linear constraints and $(n+2d)$ variables. Solving these linear programs require significantly more resources (time and memory) compared to OLS~\cite{portnoy1997gaussian}, making {\em QR} prohibitive for all but small-to-moderate sized datasets.
This need for extensive computational resources has been one of the main reasons for the relative lack of adoption of {\em QR} in emerging Big Data applications. 

\stitle{State of Art Techniques and Tools:}  Quantile Regression is a well-studied research problem. In the appendix (Section \ref{sec:related_work}), we provide a more detailed overview of the literature, but highlight a few key works here.
%An overview of literature has been provided in~\cite{?}.
% A complete overview of literature is beyond the scope of this paper.
% Existing work often modifies the original objective function to achieve a faster performance \cite{?,?}. 
% As none of these modification are widely accepted to our best knowledge, in this paper, we are only interested in the classical quantile regression problem.
%While a bulk of existing work~\cite{?,?} modifies the original objective function to achieve a faster performance,
%in this paper we focus on the original problem, and the exact quantile regression techniques which find the optimal solution.
% In this paper, we only focus on exact quantile regression techniques which truly finds the optimal solution.

Two prominent classes of exact techniques to solve the {\em QR} problem are {\em exterior-point based}~\cite{barrodale1973improved} and {\em interior-point based}~\cite{portnoy1997gaussian} approaches.
Barrodale and Robert ({\em BR}) proposed a simplex-based  exterior point technique~\cite{barrodale1973improved}, which moves from one exterior point (i.e., a ``corner'' of the feasible polytope defined by the linear program) to another exterior point, in the direction of steepest gradient descent. 
The core idea of {\em BR} originates from Edgeworth's bi-variate weighted median based 1888 approach \cite{edgeworth1888xxii}. The run time of this algorithm is  $O(n^2)$, which is still the state of the art algorithm for two dimensions (i.e., $d =2$). 
% As the de-facto exterior-point solution for QR, this approach has been extensively implemented across different platforms~\cite{quantreg}.
% There has been extensive developments based on this work \cite{?,?,?}.
% However, as we shall demonstrate in our experiments in \S~\ref{sec:experiments}, 
All the exterior points based approaches are only suitable for small problem instances with around $1000$ points and few attributes \cite{portnoy1997gaussian}.
Portnoy and Koenkar later proposed a primal-dual interior point method ({\em IPM})  which finds the optimal solution by minimizing the difference between primal and dual objective cost~\cite{portnoy1997gaussian}.
This method is reasonably fast for larger input sizes in practice, but worst-case theoretical runtime complexity is $O(n^{3.5}\log {1/\epsilon})$ where $\epsilon$ is the desired accuracy \cite{wright1997primal}. 
Moreover,  {\em BR} and {\em IPM} are extremely memory hungry, hence not suitable for big data problem instances under resource constraints.

\section{Related work and discussion on state of the art}\label{sec:related_work}
The idea of quantile regression was introduced even before {\em OLS}  regression.
Around 1757, Boscovich proposed the idea of fitting a line for 2-dimensional data
by minimizing the sum of absolute residuals under the assumption of mean of residuals has to be zero.
Laplace gave an algebraic formulation of the problem in his Methode de Situation in 1789.
In 1809 Gauss suggested to remove zero mean residual constraint.
Later in 1823, he also proposed  least square criterion (i.e., {\em OLS}). 
As {\em OLS} has more analytical and computational simplicity, it has been always popular.
However, a criterion is a choice and there are different cases where one criterion will outperform others.
In 1888, Edgeworth proposed a geometry based solution for the bi-variate median regression problem \cite{bloomfield1983least}.  

In the 20th century, there has been extensive work on quantile regression which helped a wide range of applications. 
In 1974, Barrodale and Roberts \cite{barrodale1973improved} used simplex technique to solve median regression problem as a bounded dual problem. 
Bloomfield and Steiger \cite{bloomfield1983least} explored   the simplex technique for median regression in depth and suggested exploring a normalized steepest edge direction instead of steepest edge direction.

Before 1987, all the quantile regression techniques were focusing on median regression Koenker and d’Orey \cite{koenker1987algorithm} generalized the criterion for any quantile which is now known as quantile regression.
In median regression,  an input point incurs a cost of absolute deviation.  
For quantile regression parameter $\tau$, Koenker and d’Orey proposed to assign $1-\tau$ and $\tau$ weight to negative and positive residuals. 
After the development of generalized quantile regression, there has been an extensive research in this area. 
We mention a few prominent work in this paper. 
The primary focus has been on the development of faster quantile regression algorithms \cite{chernozhukov2020fast}. 
Another research direction is finding  quantile regression for multiple quantiles \cite{portnoy1997gaussian, chernozhukov2020fast}. 
In 1997, Portnoy and Koenker \cite{portnoy1997gaussian} developed a primal dual interior point method where in each step the goal is to minimize the different between primal dual objective loss. 
It is extremely fast compared to the all previous methods.

In the continuous quest for faster quantile regression, either the objective criterion has often been  modified~\cite{chernozhukov2020fast, yang2013quantile} or approximations~\cite{meinshausen2006quantile,feng2015bayesian,zheng2011gradient} have been introduced . 
As none of these are widely accepted,  we only consider the original quantile regression problem in this paper. 
If we we consider the entire history of quantile regression, the most prominent two exact techniques are based on Barrodale and Roberts\cite{barrodale1973improved}, and Portnoy and Koenkar \cite{portnoy1997gaussian}. 
The latest updated implementation based on these techniques can be found in quantreg package \cite{quantreg} in the R library, maintained by Roger Koenkar which is a gold standard public library for quantile regression. 
However, these widely used and accepted implementations are quite memory hungry.
Moreover, the run time complexity of these implementations are still far behind to tackle the challenges of modern big data applications.
So it is extremely important to develop techniques that can overcome these ever increasing challenges.

\subsection{Existing State-of-Art Exact Approaches}\label{subsec:state-of-art}
In this subsection, we describe a few properties of {\em QR}, connection to linear programming, and existing exact techniques to solve  problem. 

There has been  extensive research conducted on  {\em QR} over the last two hundred years~\cite{bloomfield1983least}.
Wagner \cite{wagner1959linear} identified the connection between linear programming and {\em $\ell_1$-regression}.
A reparameterization technique\cite{koenker1978regression}  is used to convert the {\em QR} problem into a linear programming problem.
The linear constraints of a linear programming problem create a convex polytope.
Solutions to the {\em QR} problem are based on utilizing the exterior (corner) points or interior points of this convex polytope.

We give a brief overview of prior algorithms for {\em QR}. The focus of {\em QR} research has been primarily on how to make the algorithms faster, with the eventual goal of trying to make it an alternative to {\em OLS} regression.
% \textcolor{red}{Gautam: OLS needs to be defined.}  
There have been two broad categories of techniques, exact techniques (where the objective function is minimized) and approximate/heuristic techniques. There is a wide body of techniques that belong to the latter category, e.g., modifying the original objective function to achieve a faster performance~\cite{chernozhukov2020fast, yang2013quantile}, as well as other approximation approaches ~\cite{meinshausen2006quantile,feng2015bayesian,zheng2011gradient}. However, since our focus in this paper is to consider regression as a robust and trustworthy technique, we are only interested in exact approaches. 

Prior research on exact quantile regression techniques can be divided into two categories: {\em exterior point based approaches} and {\em interior point based approaches}.
% \textcolor{red}{Gautam: can you give a line or two giving some intuition why these are called interior and exterior point methods? Where is the connection with Linear Programming so far?} 
% One of the earlier exterior point method based exact approaches is Barrodale and Roberts (BR)\cite{barrodale1973improved}. 
In 1888, Edgeworth designed an exterior point approach~\cite{edgeworth1888xxii} to solve 2-dimensional {\em $\ell_1$}-regression.
A modern implementation of Edgeworth's technique that utilizes a linear-time weighted median finding algorithm as subroutine can solve 2-dimensional {\em QR} in $O(n^2)$ time. 
Barrodale and Roberts ({\em BR})~\cite{barrodale1973improved} designed a simplex algorithm-based approach for any dimension that starts from an exterior point and moves in the direction of the steepest gradient descent of the objective function.
This algorithm runs in $O(n^2)$ time for 2-dimensional {\em QR}.
However, these types of approaches are not very scalable for large datasets.

% The {\em quantreg}\cite{quantreg} package in $R$ includes the implementation of BR in their package, which we named {\em QR-BR}.
% % Unfortunately, (as we will extensively investigate later in our experiments section),
% Unfortunately, {\em QR-BR} is extremely slow and memory inefficient for large datasets. 

% \textcolor{red}{Gautam: You say that this is one of the earliest. So it is not the latest? If so, why are we not discussing the latest state-of-art? Also, important to mention that this is part of packages such as R}.
% \textcolor{red}{Gautam: Readers will be lost here. What is simplex algorithm? What is feasible point? What is neighboring feasible point? So far, you have not mentioned linear programming at all, nor a geometric description of the problem.}
% The idea is  essentially build upon  Edgeworth's paper from 1888 \cite{edgeworth1888xxii} regarding $l_1$ regression.
% \textcolor{red}{Gautam: What is $l_1$ regression?} 
% From a feasible point move along the direction of the steepest gradient descent until the minimum along the direction is found. 
% From one minimum move to another minimum until there is no reduction in objective loss. 
% \textcolor{red}{Gautam: What does all this mean? ``Feasible point'', ``from one minimum, move to another minimum"?} 

%For large datasets, interior point based solutions are more appropriate.
Portnoy and Koenkar \cite{portnoy1997gaussian} introduced a primal-dual interior point method ({\em IPM}) solution for {\em QR}. 
In this approach, the {\em QR} problem is solved using primal and dual space both simultaneously.
{\em IPM} adds a barrier function in the objective function, which helps explore the the solutions in interior points. 
The algorithm stops when the primal and dual objective difference is smaller than a given small threshold $\epsilon$.
% The {\em quantreg} package includes the latest {\em IPM} based implementation, which we call {\em QR-IPM}.
The worst case run time complexity for {\em IPM} is $O(n^{3.5}\log{1/\epsilon})$ \cite{wright1997primal}.
Although {\em IPM} is most widely used {\em QR} technique in practice, it is not yet ready to handle big data.

The detailed related works can be found in Appendix \ref{sec:related_work}. 

% The {\em quantreg} package includes the latest {\em IPM} based implementation, which we call {\em QR-IPM}.
% \textcolor{red}{Gautam: Again, I think the whole business of primal, dual, etc, would be better understood if you first describe the geometry of the problem, and its connection with linear programming.}
% In each iteration, the goal is to move from one interior point of the feasible area to another interior point such that the difference between primal and dual problems' objective values is minimized. 
% The algorithm stops when the difference is below a given threshold. 
% For a small threshold value, the solution is really close to the optimal solution. 
% \textcolor{red}{Gautam: Is this the latest state-of-art interior point method? If it is, say so. Also mention that this algorithm is available in the R package}.
% Although, {\em QR-IPM} is quite fast, it is extremely memory hungry. 
% For example, in our experiments with large datasets the {\em QR-IPM} often results in the operating system thrashing. 
% Our experimental evaluation will demonstrate the shortcomings of {\em QR-BR} and {\em QR-IPM} in detail.

% \section{Arrangements and its \texorpdfstring{$k$}-th level}\label{subsec:arrangements-ksets}
% One of the contributions of our paper is the connection of computational geometry notions of {\em arrangements} and {$k$-sets} with the seemingly unrelated problem of quantile regression.  We briefly review these concepts below.

\section{Details of k-set based two dimensional QR problem}
\subsection{\neighborBaseline : Algorithm Based on Neighbor Exploration}\label{subsec:update-neighbor-approach-2d}

In  this subsection, we introduce \neighborBaseline algorithm that utilizes the \un. 

Exploring the vertices in the complete skeleton graph presents us with an interesting algorithm to obtain the optimal {\em QR} line.
Let us consider for simplicity that a {\sc GetNeighbors} oracle exists that can quickly provide us with the neighbors of a given vertex.
We can start from any of the $\mathcal{O}(n^2)$ vertices and explore neighboring vertices around it in a Breadth First Search (BFS) manner while updating the aggregate entities and objective cost.
While for simplicity of understanding, we choose BFS for exploration, it does not change the analysis that follows.
We stop when no new neighboring vertices with a lower optimization value are found. 
The optimal QR hyperplane is the vertex with the smallest optimization function among the explored intersection points.
% Note that one can go over the points differently than BFS.
% The analysis and the following argument below would still hold. 

% A naive Oracle can be constructed by computing all the intersections points beforehand.
Note that a naive way to build a {\sc GetNeighbors} Oracle is to compute the $\mathcal{O}(n^2)$ vertices of the complete skeleton graph beforehand. 
Computing and storing intersection points consumes $\mathcal{O}(n^2)$ space and $\mathcal{O}(n^2)$ time.
Often the $\mathcal{O}(n^2)$ space is prohibitive in practice.
As the dimensions grow beyond $2$, this neighborhood-based approach is prohibitive as there are a total of $\mathcal{O}(n^d)$ points.
Exploring these intersection points is far more expensive than the {\em IPM} for larger dimensions.

This algorithm provides an approach that explores the neighbors of vertices until it reaches the optimal solution.
While this approach produces a similar complexity as the exterior point method for two dimensions, it creates an intuition for our optimized 2D algorithm.
We present a more efficient approach for two dimensions later in this section.

\subsection{\ksetBaseline: \texorpdfstring{$k$}\ -Set Based Algorithm}\label{subsec:ksetBaseline-2d}
In this subsection, we introduce \ksetBaseline algorithm that utilizes the computational geometry concept of $k$-set. 

%The seemingly disconnected concept of arrangement is connected to the problem of {\em QR} through the concept of $k$-sets.
Recall from Section~\ref{sec:geom-map} that a $k$-set is a subset containing $k$ points that can be separated from the rest of the $n-k$ points by a hyperplane.
$k$-sets have often been used in various computational geometry settings to solve numerous problems.
% In this subsection, we connect the seemingly disconnected problem of QR to the concept of $k$-sets.

% The seemingly disconnected concept of arrangement is connected to the problem of {\em QR} through the concept of $k$-sets.
An interesting observation that can be made on the optimal solution {\em QR} hyperplane is that it separates the $n$ points into $\tau n$ and $(1-\tau)n$ points \cite{portnoy1997gaussian}.
While there are many hyperplanes that separate $n$ points into two parts (with $\tau n$ and $(1-\tau)n$ points), we are interested in the specific hyperplane that provides the lowest optimization score.
As $k$-sets separate the points into $k$ and $n-k$ points, the optimal hyperplane must be one of the $k$-set separating hyperplanes such that $k=\tau n$.
The enumeration of all $k$-sets can also be viewed as a walk along the $k$-level of an arrangement.
% such that $k$ vertices lie above the vertices belonging to the curve in the dual space.
Hence, the optimal hyperplane can be found along the vertices encountered during the walk on the $k$-level of the arrangement.

% For a given value of $\tau$ and $n$, the optimal hyperplane must separate the points into $\tau n$ and $(1-\tau)n$ points.
% Hence, the optimal hyperplane that partitions the points must be a hyperplane corresponding to a $k$-set with $k=(1-\tau) n$.
% \suraj{Add an example of $k$-set from example}

The {\em QR} problem in 2 dimensions can utilize this property (the result of Theorem \ref{thm:opt_k_point}).
The optimal hyperplane $\mathcal{H}^*$ in primal space is represented by a point $\mathcal{F}(\mathcal{H}^*)$ in dual space.
This point is one of the vertices in the complete skeleton graph corresponding to the $k$-level arrangement in the dual space.
One approach to solving this problem would be to enumerate the vertices corresponding to the $k$-level of the arrangement and, for each vertex, compute the value of the optimization function.
% The best known upper bound on the total number of $k$-sets~\cite{dey1997improved} for a given value of $k$ is proved to be $\mathcal{O}(nk^{1/3})$.
% Computing the points in the $k$-level of the arrangement using \randalg\cite{chan1999remarks} takes $\mathcal{O}(n\log{(n)}) + nk^{1/3})$ expected time.% Corollary 4.4
% We will first show that these results in themselves without the use of \un would result in an inefficient algorithm.
% and then we present our $2$ dimensional QR algorithm.

% \subsection{$k$-set exploration approach without \un}

We first present \ksetBaseline algorithm that makes use of the concept of $k$-sets  without the use of aggregate entities to solve the problem. 
This approach has to compute the value (error) of the optimization function for a vertex that corresponds to a dual point $p$ in the $k$-level arrangement.
Computing the optimization function for each of these vertices involves aggregating the error contribution of each of the $n$ points in the primal space using the line $\mathcal{F}^{-1}(p)$.
Note that each of the computations consumes $\mathcal{O}(n)$ time.
For each of the vertices corresponding to the dual points in the $k$-level arrangement, the naive error computation is repeated, and the optimal line $l^*$ corresponds to that line $\mathcal{F}^{-1}(p)$ with the least error.

\stitle{Time complexity}:
For a given vertex $p$ in the $k$-level arrangement, a naive approach to compute the error without making use of \un  takes a total of $\mathcal{O}(n)$ time.
The upper bound on the intersection points in $k$-level of an arrangement is bounded by $\mathcal{O}(nk^{1/3})$.
Hence, the total time taken for the naive approach is $\mathcal{O}(n^2k^{1/3})$ for a given $\tau$ ($k=\tau n$).
As $k=n/2$ when $\tau=1/2$, the worst case time complexity is $\mathcal{O}(n^{7/3})$.

\subsection{Illustration of \un}\label{subsec:illustrate-UN}
% We will bring out the limitations of this expensive computational approach when we introduce our QR algorithms for 2-dimension.
In this subsection, we illustrate how we can reduce the computational cost while moving from one vertex to a neighboring vertex of the complete skeleton of the arrangement by  maintaining only a few aggregate values. For this illustration of \un,  we use our running example from Table~\ref{tbl:running_ex} with $\tau=0.5$.

% From Theorem~\ref{thm:opt_k_point}, the optimal {\em QR} hyperplane  lies on $d$ input points.
% As a result, the optimal solution is a point which lies on the intersection of $d$ dual hyperplanes in dual space.
% \suraj{Update Figure 1 with 2 neighbouring points and update aggregate values from table 1}
% To explain the application of Theorem~\ref{thm:constant_update}, let us walk you through with an example.

\begin{wrapfigure}{r}{0.48\textwidth}
    \centering
    \vspace{-8mm}
    \includegraphics[width=0.48\textwidth]{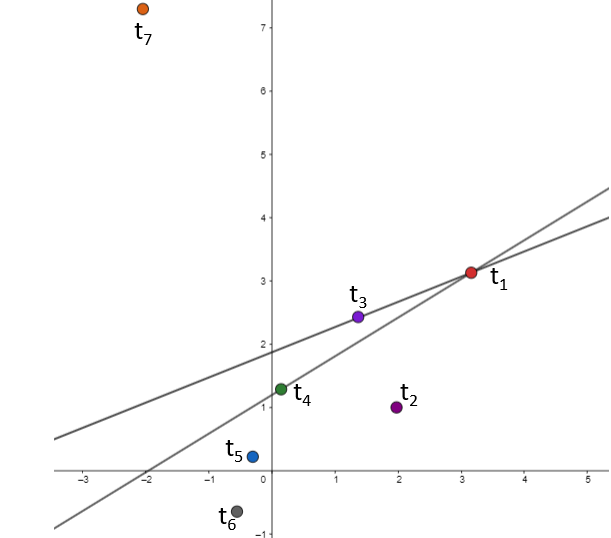}
    \vspace{-5mm}
        \caption{Update Operation}
        \label{fig:orderd_update}
     \vspace{-2mm}
\end{wrapfigure}

As explained in Section \ref{sec:preliminaries}, a vertex in the complete skeleton graph (point in dual space) refers to a {\em QR} hyperplane in primal space.
The optimal solution lies in one of these vertices in the complete skeleton graph.
Let $S$ be a complete skeleton graph constructed from Figure \ref{fig:dual_space} and $P_{(i,j)}$ be any vertex constructed from the intersection of dual line $T_i$ and $T_j$. 
% As shown Figure \ref{fig:dual_space}, $P_{(1,3)}$ and  $P_{(1,4)}$ are neighbors which will be used in our illustration.
$P_{(1,3)}$ and $P_{(1,4)}$ are neighbors in the arrangement which will be used in our illustration.
Figure \ref{fig:orderd_update} shows the primal representation of  $P_{(1,3)}$ and  $P_{(1,4)}$ where each represents a line.
% For this illustration, a line divides the plane into two halves. 
From the definition of residual set in Section \ref{sec:preliminaries},  $I^+=\{1, 3, 4, 7\}$ and  $I^-=\{2, 5, 6\}$ for $P_{(1,4)}$.
We calculate the sum along $X$-axis and $Y$-axis for these two group of points and call these  aggregate values at $P_{(1,4)}$.
Given these aggregate values for $P_{(1,4)}$,  we would like to answer the following questions as steps of  our demonstration: (i) How to calculate the objective values at $P_{(1,4)}$? (ii) How to calculate the aggregate values at $P_{(1,3)}$ from $P_{(1,4)}$ in $\mathcal{O}(1)$? and (iii) If aggregate values are known for $P_{(1,3)}$, how to calculate the objective value at $P_{(1,3)}$?
% \begin{enumerate}[leftmargin=*]
%     % \item How to calculate the aggregate values at $P_{(1,4)}$?
%     \item How to calculate the objective value at $P_{(1,4)}$?
%     \item How to calculate the aggregate values at $P_{(1,3)}$ from $P_{(1,4)}$ in $O(1)$?
%     \item If aggregate values are known for $P_{(1,3)}$, how to calculate the objective value at $P_{(1,3)}$?
% \end{enumerate}

% \begin{figure}[t]
%     \centering
%     % \includegraphics[width=0.35\textwidth]{figures/orderd_update.PNG}
%     \includegraphics[width=0.3\textwidth]{figures/OracleD.PNG}
%     \caption{Update Operation}
%     \label{fig:orderd_update}
%     \vspace{-5mm}
% \end{figure}

As shown in Figure \ref{fig:orderd_update}, $P_{(1,4)}$  passes through input points $t_1$ and $t_4$ in primal space and its parameter vector is $\beta = [1.19, 0.71]$. 
For all $i \in I^+$, let  $\sum x_i$, $\sum y_i$ be the sum along X and Y axis respectively.
Similarly, for all $j \in I^-$, let  $\sum x_j$ , and $\sum y_j$  be the sum along X and Y axis respectively.
% All of these aggregate values can be easily calculated for $P_{(1,4)}$ using the items in $I^+$ and $I^-$.
% Let  $\sum^+ x$, $\sum^+ y$ be the sum along X and Y axis for items in $I^+$, and  $\sum^- x$, $\sum^- y$ be the sum along X and Y axis for items in $I^-$.
Given the aggregate values for $P_{(1,4)}$, the objective cost $\mathcal{L}$ at $P_{(1,4)}$ can be calculated using following process.

\vspace{-3mm}
\begin{align*}
\sum {r_i}  &= (\sum{y_i}-\sum{x_i} \times \beta[1]-\beta[0] \times |I^+|) 
            % &= (14.147-2.628\time 0.71 -1.19\times 4) \\
            = 7.52 \\
\sum {r_j}  &= (\sum{x_j}\times \beta[1]+\beta[0]\times |I^-| -\sum{y_j}) 
            % &= (1.12\times 0.71 + 1.19\times 3 - 0.572) \\
            = 3.79 \\
\mathcal{L} &= 0.5\times \sum{r_i}  + (1-0.5)\times \sum{r_j}
            = 5.65
\end{align*}
\vspace{-3mm}

Now, we will try to answer how we will update the aggregate values at $P_{(1,3)}$ from $P_{(1,4)}$. 
As shown in Figure \ref{fig:orderd_update},  only one input point $t_4$ changes its residual sign from positive to negative in $P_{(1,4)}$ to $P_{(1,3)}$ transition. 
For $P_{(1,3)}$, $I^+=\{1,3, 7\}$ and $I^-=\{2, 4, 5, 6\}$. 
If we want to calculate the aggregate values at $P_{(1,3)}$ from  $P_{(1,4)}$, the information of input point $t4$ can be utilized in the following fashion. 
\vspace{-3mm}
\begin{align*}
\sum{y_i} &= \sum{y_i} -   t_4.y = 12.86, \sum{y_j} &= \sum{y_j} +   t_4.y = 1.86\\
\sum{x_i} &= \sum{x_i} - t_4.x = 2.47, \sum{x_j} &= \sum{x_j} +  t_4.x = 1.27
\end{align*}
\vspace{-3mm}

Earlier, we have shown how to calculate objective cost from aggregate values at $P_{(1,4)}$. 
Similarly, the objective cost for $P_{(1,3)}$ can be calculated using the aggregate values. 
If the aggregate values at $P_{(1,4)}$ are known, the aggregate values and objective cost at $P_{(1,3)}$ can be calculated in $\mathcal{O}(1)$ time and space.
Although our illustration is in 2-dimension, our \un works in any dimension.

\input{sections/appendix_proofs}

\input{sections/appendix_algos}

%% file: sections/appendix_proofs.tex
\section{Proofs of Theorems and Lemmas} \label{sec:lemmaApp}
\leavevmode\newline

\noindent
{\bf Theorem~\ref{thm:constant_update}}.
{\em 
% \ref{thm:qrreg2d-rand-proof}
Given the aggregate values $\forall_{1\leq m \leq d}\sum_{i \in I^+} X_i[m]$, $\sum_{i \in I^+} y_i$, $\sum_{j \in I^-} y_j$, $\forall_{1\leq m \leq d}\ \sum_{j \in I^-} X_j[m]$ for a vertex in the complete skeleton graph, the aggregate values can be updated in $\mathcal{O}(d)$ time when we move to a neighboring vertex.
}
\begin{proof}
Any vertex in the complete skeleton graph represents a hyperplane in the primal space which divides the set of points in $\mathcal{D}$ into two sets, $I^+$ and $I^-$.
When we move from a vertex to its neighboring vertex, the sets $I^+$ and $I^-$ change in one of three ways,
\begin{itemize}
    \item A point in primal space moves from above the hyperplane to below the hyperplane i.e., a point moves from $I^+$ to $I^-$.
    \item A point in primal space moves from below the hyperplane to above the hyperplane i.e., a point moves from $I^-$ to $I^+$.
    \item The hyperplane moves such that a point from above is exchanged with a point from below. 
    In such a case, the number of points above the hyperplane remains the same, i.e., a point from $I^+$ is swapped with a point in $I^-$. 
    Note that the $k$ in this case may not be $\tau(1-n)$.
\end{itemize}

For each of the three cases, we prove that the aggregate values can be updated in $\mathcal{O}(d)$ time.
For the rest of this proof, let the new sets after the update be represented by $U^+$ and $U^-$.

Let us consider the first case, where the hyperplane moves such that a point moves from the set $I^+$ to $I^-$, i.e., a point in primal space which was above the hyperplane (vertex) now lies below the neighboring hyperplane (neighbor vertex).
Let $X_t$ be the point that is involved in the transition.
As we know the details of point $X_t$, we can remove the contribution of $X_t$ towards $I^+$ and add the contribution to aggregates entities of $I^-$.
The formulae for the update are as below,
\begin{align*}
    \sum_{i \in U^+} y_i &= \sum_{i \in I^+} y_i - y_t\\
    \forall_{1\leq m \leq d}\ \ \ \sum_{i \in U^+} X_i[m] &= \sum_{i \in I^+} X_i[m] - X_t[m]\\
    \sum_{i \in U^-} y_i &= \sum_{i \in I^-} y_i + y_t\\
    \forall_{1\leq m \leq d}\ \ \ \sum_{i \in U^-} X_i[m] &= \sum_{i \in I^-} X_i[m] + X_t[m]
\end{align*}

As there are $2d+2$ equations, each of which takes $\mathcal{O}(1)$ time to update, the total time taken to update the aggregate values is $\mathcal{O}(d)$.

The second case, where a point in primal space has moved from below the hyperplane (vertex) to above the neighboring hyperplane (neighbor vertex), can be updated similarly.
Let $X_t$ be the point that is involved in the transition.
The  formulae for the update operation for the second case are,

\begin{align*}
    \sum_{i \in U^+} y_i &= \sum_{i \in I^+} y_i + y_t\\
    \forall_{1\leq m \leq d}\ \ \ \sum_{i \in U^+} X_i[m] &= \sum_{i \in I^+} X_i[m] + X_t[m]\\
    \sum_{i \in U^-} y_i &= \sum_{i \in I^-} y_i - y_t\\
    \forall_{1\leq m \leq d}\ \ \ \sum_{i \in U^-} X_i[m] &= \sum_{i \in I^-} X_i[m] - X_t[m]
\end{align*}
With $2d+2$ equations, each of which take $\mathcal{O}(1)$ the overall time complexity is $\mathcal{O}(d)$ time.

Let us consider the third case, where the hyperplane moves such that a point from above is exchanged with a point from below.
In this case, a point from $I^+$ is swapped with a point in $I^-$.
Let $X_s$ be the point that is moved from $I^+$ to $I^-$, and $X_t$ be the point moved from $I^-$ to $I^+$.
As we know the details of points $X_s$ and $X_t$, we can remove the contribution of $X_s$ towards $I^+$ and add the contribution of $X_t$ to it.
We perform vice verse operation to $I^-$.
The formulae for the update are as below,
\begin{align*}
    \sum_{i \in U^+} y_i &= \sum_{i \in I^+} y_i + y_t - y_s\\
    \forall_{1\leq m \leq d}\ \ \ \sum_{i \in U^+} X_i[m] &= \sum_{i \in I^+} X_i[m] + X_t[m] - X_s[m]\\
    \sum_{i \in U^-} y_i &= \sum_{i \in I^-} y_i - y_t + y_s\\
    \forall_{1\leq m \leq d}\ \ \ \sum_{i \in U^-} X_i[m] &= \sum_{i \in I^-} X_i[m] - X_t[m] + X_s[m]
\end{align*}
With a total of $2d+2$ equations, each of which take $\mathcal{O}(1)$, the overall time complexity is $\mathcal{O}(d)$ time.
Hence, proved.
\end{proof}

% \subsection{Proof of Lemma \ref{thm:qrreg2d-proof}}
\vspace{2mm}\noindent{\bf Lemma~\ref{thm:qrreg2d-proof}}.
% \label{thm:qrreg2d-proof}
{\em
\twodAlg has a time complexity of $\mathcal{O}(n^{\frac{4}{3}}\log^{1+a}{n})$, where $a >0$ is an arbitrarily small constant.
}

\begin{proof}
For the first $k$-set obtained through the enumeration, the optimization function and aggregate values need to be calculated by a linear scan over the points which takes $\mathcal{O}(n)$ time.
% The $2$ dimensional algorithm is given in \ref{alg:2d-algo}.
Updating the optimization function value and aggregate values as we explore neighboring $k$-set takes $\mathcal{O}(1)$ time.
The points in the $k$-level of the arrangement can be computed using  a sweep Line algorithm \cite{edelsbrunner1986constructing, chan2001dynamic, chan1999remarks} in $O(n\log{m} + m\log^{1+a}{n})$ where $m$ is the number of $k$-set and $a >0$ is an arbitrarily small constant.
The upper bound on total number of $k$-sets is given by Dey~\cite{dey1997improved}, $\mathcal{O}(nk^{1/3})$.
This brings the overall time taken to $\mathcal{O}(nk^{1/3}\log^{1+a}{n})$.
As $k$ is a percentage of $n$, the overall time complexity $\mathcal{O}(n^{4/3}\log^{1+a}{n})$.
Hence, proved.
\end{proof}

% \subsection{Proof of Lemma \ref{thm:qrreg2d-rand-proof}}

\vspace{2mm}\noindent{\bf Lemma~\ref{thm:qrreg2d-rand-proof}}.
{\em 
% \label{thm:qrreg2d-rand-proof}
\twodAlg with randomized $k$-set enumeration~\cite{chan1999remarks} has an expected time complexity of
% Algorithm~\ref{alg:2d-algo} finds the optimal solution for the 2 dimensional {\em QR} problem in 
$\mathcal{O}(n^{\frac{4}{3}})$.
}
\begin{proof}

This proof follows along similar lines to the proof of Lemma \ref{thm:qrreg2d-proof}. 
However, if we use a randomized incremental algorithm instead of the deterministic algorithm, the enumeration can be done in $\mathcal{O}(n\log{(n)} + nk^{1/3})$ expected time (Corollary 4.4~\cite{chan1999remarks}). 
The points in the $k$-level of the arrangement can be computed using \randalg\cite{chan1999remarks} in $\mathcal{O}(n\log{(n)} + nk^{1/3})$ time (Corollary 4.4). 
This brings the overall time taken to  $\mathcal{O}(n^{4/3})$.
\end{proof}

\vspace{2mm}\noindent{\bf Lemma~\ref{lem3}.}
{\em \ConsSampl is an unbiased sampler for $S_{in}$.}

\begin{proof}
    Let $S_{in}^{(i)}$ be the number of intersections in $S_{in}$ that involve $\ell_i$.
    The probability of selecting each line $\ell_i$ is $$Pr(\ell_i)= \frac{|S_{in}^{(i)}|}{2|S_{in}|}$$
    After selecting a line $\ell_i$, the probability that a specific intersection $P_j$ involving it is $\frac{1}{|S_{in}^{(i)}|}$.
    Now, for an intersection $P_j\in S_{in}$, let $\ell_i$ and $\ell_k$ be the dual lines that involve it.
    The probability of sampling $P_j$ is 
    \begin{align*}
        Pr(P_j)&= \frac{|S_{in}^{(i)}|}{2|S_{in}|}\times \frac{1}{|S_{in}^{(i)}|} + \frac{|S_{in}^{(k)}|}{2|S_{in}|}\times \frac{1}{|S_{in}^{(k)}|} \\
        &= \frac{1}{|S_{in}|}
    \end{align*}
\end{proof}

\vspace{2mm}\noindent{\bf Theorem~\ref{thm:expected-runtime-2d-zpp}}
{\em
The expected time complexity of \randQR in two dimensions is $\mathcal{O}(n\log^2{(n)})$.
}
\begin{proof}
    % Theorem~\ref{thm:expected-runtime-general-zpp} gives us a base to prove our current theorem.
    % To use the result from Theorem~\ref{thm:expected-runtime-general-zpp}, we need to compute the time complexity of (i) \SplitOrc, and (ii) sampling strategy.
    \randQR algorithm relies on two important functions - \SplitOrc and \ConsSampl.
    For this proof, we first analyze the running time of these two approaches and in a final step use these to prove the run time of \randQR.

    \stitle{\SplitOrc time complexity}: The \SplitOrc for two dimensions has two main parts (i) Binary Search routine, (ii) Update $R$.
    The binary search routine starts from an array of intersections with length $n$.
    In each step, the optimization function value is computed for the mid-point, which consumes linear time ($\mathcal{O}(n)$).
    As binary search finds the answer in $\mathcal{O}(\log{(n)})$ steps, the total time consumed by the binary search is $\mathcal{O}(n\log{(n)})$.

    Once the optimum in the plane $z_i=v$ is computed, the intersections of the corresponding dual line with other $n-1$ lines are obtained in linear time.
    Computing the optimization value for the neighboring intersections can be performed in ($\mathcal{O}(n)$) time.
    Thus the total time complexity of \SplitOrc is $\mathcal{O}(n\log{(n)})$.

    \stitle{Time complexity of \ConsSampl}: 
    The first step in the weighted sampling strategy is to determine the weights of the dual lines.
    Our approach internally uses the {\em inversion counting problem} to obtain the weights.
    The inversion counting problem is well known,  and runs in time $\mathcal{O}(n\log{(n)})$.
    These weights are then used to sample the first dual line.
    In linear time, the intersections of the sampled line with all the other $n-1$ lines are obtained.
    One of these lines that satisfy $R$ is sampled.
    Thus, the time complexity is dominated by the inversion counting problem.

    The sum of the time complexities of the divide and conquer steps of our algorithm is $\mathcal{O}(n\log{(n)})$.
    Let the time taken by this single divide-and-conquer step be denoted entity by $T(n, 2)$ (where $2$ represents the number of dimensions).

    Our proof follows along similar lines as the {\em Randomized Quick-Sort} algorithm.
    Let the interval contain $N=n^2$ vertices of $S$.
    The recurrence for our problem is given below,
    \begin{align*}
        \mathbb{E}[F(n, 2, N)]&=\frac{1}{N}\overset{N}{\underset{j=1}{\sum}}\mathbb{E}[\max{(F(n, 2, N- j), F(n, 2, N -(N-j)))}]+\Theta(T(n, 2))
    \end{align*}
    where $N-j$ represents the removal of $j$ vertices from the range.
    Assume that the $\mathbb{E}[F(n, 2, R)]=\mathcal{O}(T(n, 2)\log{(n)})$ (i.e. $\mathbb{E}[F(n, 2, N)]=c_0\log{(n)} T(n, 2)$).
    Note that one of the $N$ recursive calls consumes $\mathcal{O}(T(n, 2)$ time as it corresponds to the optimal hyperplane. 
    \begin{align*}
        c_0\ \log{(n)}\ T(n ,2) &\stackrel{?}{<} c_1\:T(n, 2) + \frac{1}{N} \overset{N-1}{\underset{j=1}{\sum}} c_0\:\log{(n)}\ T(n, 2)\\
        c_0\ \log{(n)}\ T(n, 2) &\stackrel{?}{<} c_1\: T(n, 2) +  \frac{N-1}{N} c_0\:\log{(n)}\ T(n, 2)
        % c_0\ \log{(n)}\ T(n^d) &\stackrel{?}{<} c_1\ T(n^d) + \frac{1}{n^d} \overset{n^d-1}{\underset{j=1}{\sum}} c_0\ \log{(n)}\ T(n^d) = c_1\ T(n^d) +  \frac{n^d-1}{n^d} c_0\ \log{(n)}\ T(n^d)
    \end{align*}
    By choosing the right value for constants $c_0$ and $c_1$, the above inequality can be satisfied.
    Thus, the expected run-time of our \randQR approach is $\mathcal{O}(\:\log{(n)}\ T(n, 2)\:)$ = $\mathcal{O}(n\log^2{(n)})$.

    % The overall time complexity of both these parts, which is termed as $T(n^2)$ in Theorem~\ref{thm:expected-runtime-general-zpp} is $\mathcal{O}(n\log{(n)})$.
    % Hence, the overall time complexity of our $\mathbb{ZPP}$ approach in two dimensions is $\mathcal{O}(T(n^2)\log{(n)})=\mathcal{O}(n\log^2{(n)})$
\end{proof}

{\sc Theorem~\ref{thm:expected-runtime-3d-zpp}}
{\em
The expected time complexity of our \randQR algorithm in higher dimensions ($d \geq 3$) is $\mathcal{O}(dn^{d-1}\log^2{(n)})$.
}

\begin{proof}
    \randQR algorithm depends on \SplitOrc and \ConsSampl. 
    We explore the time complexity analysis of these two methods before the analyze the \randQR function in $d$ dimensions.
    For this proof let $F(n, d, N)$ denote the time complexity of \randQR function in $d$ dimensions, where $N=n^d$ is the number of vertices in $S$.
    Note that all the terms used in this proof are in expected time.

    \stitle{\SplitOrc}: The \SplitOrc in $d$ dimensions calls the $d-1$ dimensional \randQR function. 
    Additionally, the $d$ hyperplanes are used to obtain $2d$ neighboring intersections (vertices).
    Each of the $\binom{d}{d-1}$ hyperplanes gives us a line in the $d$ dimensional space.
    Each of these lines is then used to compute the $n-d$ intersections (vertices) with the remaining planes.
    The total time complexity to compute the intersections is $2nd+d^2=\mathcal{O}(nd)$.
    Additionally, the optimization value for each of the $2d$ neighboring vertices is computed.
    As each computation consumes $nd$ time, the total time taken for these computations is $\mathcal{O}(nd^2)$.
    Hence, the total time complexity of \SplitOrc is $F(n, d-1, N) + nd^2$.

    \stitle{\ConsSampl}: The \ConsSampl function uses $n$ recursive lower dimensional \ConsSampl calls  to obtain the weights of each hyperplane.
    Using these weights the first hyperplane is sampled.
    Then $n-1$ recursive calls are made to $d-2$ dimensional \ConsSampl calls  to obtain the weights of each hyperplane.
    The process of obtaining weights and sampling continues until $d$ hyperplanes are sampled which intersect at a vertex inside $R$.
    Let $S$ represent the time complexity of \ConsSampl in $i$ dimensions ($S(n, i)$).
    The recursive calls made by \ConsSampl in $d$ dimensions can be captured with the equation $\sum_{i=0}^{d-1}(n-i)S(n, d-i)$.
    For two dimensions, the time complexity of \ConsSampl is $S(n, 2)=n\log{(n)}$ .
    \begin{align*}
        S(n, d) &= \sum_{i=0}^{d-1}(n-i)S(n, d-i)\\
        S(n, d)&= nS(n,d-1) + \sum_{i=0}^{d-2}(n-1-i)S(n-1, d-1-i) < nS(n,d) + \sum_{i=0}^{d-2}(n-i)S(n, d-1-i) \\
        S(n, d) &< nS(n,d-1) + \sum_{i=0}^{d-2}(n-i)S(n, d-1-i)
    \end{align*}
    But we know that
    \begin{align*}
        S(n,d-1) = \sum_{i=0}^{d-2}(n-i)S(n, d-1-i)
    \end{align*}
    Using the above expression we get,
    \begin{align*}
        S(n, d) &< nS(n,d-1) + S(n,d-1) = (n+1)S(n, d-1) \in \mathcal{O}(n\ S(n, d-1))
    \end{align*}
    For three dimensions, using the above equation we get $S(n, 3) = \mathcal{O}(n^2\log{(n)})$.
    In $d$ dimensions, we get $S(n,d) = \mathcal{O}(n^{d-1}\log{(n)})$.
    
    As the divide-and-conquer steps are performed in succession, let us consider a term which is the sum of the time complexities of these two functions.
    Let $T(n, d)=F(n, d-1,N)+n^{d-1}\log{(n)}$ denote the term that corresponds to the sum of these two functions.

    Similar to the proof for two dimensions, we use $N=n^d$ be the vertices in $R$.
    Let $F(n, d, N)$ be the time complexity of \randQR in $d$ dimensions.
    \begin{align*}
        F(n, d, N)=T(n, d) + \sum_{i=1}^{N-1}\frac{1}{N}T(n, d)=T(n,d)+\frac{N-1}{N}T(n, d)
    \end{align*}

    Assume that each $T(n, d)$ call consumes $F(n,d,N)\log{(N})$ time.
     \begin{align*}
        c_0\ \log{(n)}\ T(n ,d) &\stackrel{?}{<} c_1\:T(n, d) + \frac{1}{N} \overset{N-1}{\underset{j=1}{\sum}} c_0\:\log{(n)}\ T(n, d)\\
        c_0\ \log{(n)}\ T(n, d) &\stackrel{?}{<} c_1\: T(n, d) +  \frac{N-1}{N} c_0\:\log{(n)}\ T(n, d)
        % c_0\ \log{(n)}\ T(n^d) &\stackrel{?}{<} c_1\ T(n^d) + \frac{1}{n^d} \overset{n^d-1}{\underset{j=1}{\sum}} c_0\ \log{(n)}\ T(n^d) = c_1\ T(n^d) +  \frac{n^d-1}{n^d} c_0\ \log{(n)}\ T(n^d)
    \end{align*}

    Note that the expectation of a product of random variables is the product of the expectation of the two random variables.
    Similar to the proof for two dimensions, by choosing right constants for $c_0$ and $c_1$ we can satisfy the above inequality.
    The overall time complexity can be written in the form,
    \begin{align*}
        F(n, d, N) &= \log{(N)}T(n, d) = d\log{(n)}T(n,d)\\
        F(n, d, N) &= d\log{(n)}(F(n, d-1,N)+n^{d-1}\log{(n)})\\
        F(n, d, N) &= d\log{(n)}F(n, d-1,N)\ +\ dn^{d-1}\log^2{(n)}
    \end{align*}
    The equation for dimension $3$ gives us $F(n, 3, N)=d\log^3{(n)}n + dn^2\log^2{(n)}=\mathcal{O}(dn^2)\log^2{(n)}$.
    Similarly, for general $d$, the term $dn^{d-1}\log^2{(n)}$ grows at a faster rate compared to $d\log{(n)}F(n, d-1,N)$.
    Hence the overall complexity is $\mathcal{O}(dn^{d-1}\log^2{(n)})$
\end{proof}

%% file: sections/appendix_algos.tex
\section{Pseudocodes}\label{sec:algorithms-appendix}

% Due to space constraints,  for some of the algorithms could not be included in the main manuscript. 
% The pseudo-code for those algorithms are presented below.

\begin{algorithm}[H]
\caption{\twodAlg}
\begin{algorithmic}[1]
\label{alg:2d-algo}
\STATE {\bf Input:} Data set $\mathcal{D}$, $\tau$
\STATE {\bf Output:} Parameters for optimal line $\mathcal{H}^*$

\STATE $k\leftarrow(1-\tau)n$
\STATE $Yp\ =\ Ym\ =\ Xp\ =\ Xm\ =\ 0$ \COMMENT{Initialize aggregate sums to 0. $Yp$: $\sum y_i$, \\ $Xp$: $\sum X_i$ for $i \in I^+$ $Ym$: $\sum y_j$, $Xm$: $\sum X_j$ for $j \in I^-$}
\STATE $OptValue \leftarrow \infty$; $\ \  OptLine \leftarrow \emptyset$
\STATE Obtain $k$-level arrangement points $\mathcal{P}$ 
\STATE Calculate the aggregate values  $Ym, Xm, Yp, Xp $ using $l$ for the first point in $\mathcal{P}$
\FOR{$p$ in $\mathcal{P}$} 
    \STATE $l \leftarrow \mathcal{F}^{-1}(p)$ \COMMENT{$y=mx+c\ \implies l \leftarrow [m ,c]$}
    % \STATE $init=True$
    % \FOR{$i=1\ $\TO$n$} 
    %     \IF{$\mathcal{D}_i[1]*l[0] + l[1] > y_i$}
    %         \STATE $Ym = Ym + y_i$
    %         \STATE $Xm = Xm + \mathcal{D}_i[1]$
    %     \ELSE 
    %         \STATE $Yp = Yp + y_i$
    %         \STATE $Xp = Xp + \mathcal{D}_i[1]$
    %     \ENDIF
    % \ENDFOR
    \STATE Find out $X_s$ and $X_t$ that swapped from $I^+$ and $I^-$ respectively
    % \STATE $Ym = Ym + y_s - y_t$
    % \STATE $Xm[1] = Xm[1] + \mathcal{D}_s[1] - \mathcal{D}_t[1]$
    % \STATE $Yp = Yp - y_s + y_t$
    % \STATE $Xp[1] = Xp[1] - \mathcal{D}_s[1] + \mathcal{D}_t[1]$
    \STATE Update $Ym$, $Xm$, $Yp$, $Xp$ using \un
    \STATE $Value \leftarrow \tau(Yp-Xp\times l[1]-l[0]\times k) + (1-\tau)(Xm\times l[1]-l[0]\times (n-k)-Ym)$
    \IF{$Value < OptValue$}
        \STATE $OptValue \leftarrow LineValue$; $\ \ \ OptLine \leftarrow l$
    \ENDIF
\ENDFOR
\STATE {\bf return} $OptLine$
\end{algorithmic}
\end{algorithm}

\begin{algorithm}[H]
\caption{\texttt{BinarySearch2D}}
\begin{algorithmic}[1]
\label{alg:binary-search-2d}
\STATE {\bf Input:} Point set $\mathcal{D}$, $\tau$, intersections $points$
\STATE {\bf Output:} Best score in plane $z_i=v$, corresponding intersection point

\STATE $low \leftarrow 0$; $high \leftarrow n$; $OptScore\leftarrow \infty$
\WHILE{$low \neq high$}
    \STATE $mid \leftarrow \frac{low+high}{2}$
    \STATE $MidScore \leftarrow $\texttt{ComputeScore}$(\mathcal{D}, \tau,\beta =points[mid])$
    \STATE $NieghborScore \leftarrow $\texttt{ComputeScore}$(\mathcal{D}, \tau,\beta =points[mid+1])$
    \IF{$MidScore > NieghborScore$} 
        \STATE $high \leftarrow mid-1$ ; $OptScore\leftarrow NieghborScore$
    \ELSE
        \STATE $low \leftarrow mid$ ; $OptScore\leftarrow MidScore$
    \ENDIF
\ENDWHILE
\STATE \textbf{return} $OptScore$, vertex in $S$ corresponding to $points[low]$
\end{algorithmic}
\end{algorithm}

\begin{algorithm}[H]
\caption{\texttt{ComputeInterval}}
\begin{algorithmic}[1]
\label{alg:compute-interval-2d}
\STATE {\bf Input:} Point set $\mathcal{D}$, $\tau$, search interval $R=(R_s, R_e)$, attribute $A'$, vertex $OptLine$ in $S$ with the least score $OptScore$
\STATE {\bf Output:} Updated search interval $R$
\STATE Compute intersections of $OptLine$ with the other $n-1$ dual lines
\STATE Let $LeftLine$ and $RightLine$ be the lines which intersect with $OptLine$ immediately to the left and to the right of $z_i=v$
\STATE $score_1=$ \texttt{ComputeScore}(\texttt{Intersect}($OptLine, LeftLine$)) 
\STATE $score_2=$ \texttt{ComputeScore}(\texttt{Intersect}
($OptLine, RightLine$))
\IF{$score_1 < OptScore$}
    \STATE $R\leftarrow (R_s, v)$
\ELSIF{$score_2 < OptScore$}
    \STATE $R\leftarrow (v, R_e)$
\ELSE
    \STATE $R\leftarrow [v, v]$
\ENDIF
\STATE \textbf{return} $R$
\end{algorithmic}
\end{algorithm}

\begin{algorithm}[H]
\caption{\randQR approach}
\begin{algorithmic}[1]
\label{alg:generic-zpp-algo}
\STATE {\bf Input:} Point set $\mathcal{D}$, $\tau$
\STATE {\bf Output:} Parameters for optimal hyperplane $\mathcal{H}^*$

\STATE Choose an arbitrary variable $A'$ and set $R \leftarrow (-\infty, \infty)$
\WHILE{$true$}
    \STATE $P_j\leftarrow$\ConsSamplFn($\mathcal{F}(\mathcal{D})$, $R$, $A'$)
    % Sample vertex $P_j$ from dual of $\mathcal{D}$ which satisfies $R$
    \STATE $result \leftarrow $\SplitFn($\mathcal{D}, \tau, P_j[A']$)
    \STATE \textbf{if} $P_j$ is optimal \textbf{then} \ \ \textbf{return} $P_j$
    \STATE \textbf{else if} interval $(R_s, P_j[A'])$ contains optimal \textbf{then} $R \leftarrow (R_s, P_j[A'])$
    \STATE \textbf{else} $R \leftarrow (P_j[A'], R_e)$
    % \IF{$P_j$ is optimal}
    %     \STATE \textbf{return} $P_j$
    % \ELSIF{interval $(R_s, P_j[A'])$ contains optimal}
    %     \STATE $R \leftarrow (R_s, P_j[A'])$
    % \ELSE
    % \STATE $R \leftarrow (P_j[A'], R_e)$
    % \ENDIF
\ENDWHILE
\end{algorithmic}
\end{algorithm}
\setlength{\textfloatsep}{2pt}

\begin{algorithm}[H]
\caption{\SplitOrc for two dimensions}
\begin{algorithmic}[1]
\label{alg:split-pointers-2d}
\STATE {\bf Input:} Point set $\mathcal{D}$, $\tau$, vertex $P_j$ in $S$, search interval $R$, variable $A'$
\STATE {\bf Output:} Updated search interval $R$

\STATE $v\leftarrow P_j[A']$
% \STATE $n \leftarrow$ Number of points in $\mathcal{D}$
\STATE $points \leftarrow $ \texttt{Intersections}($z_i=v$, $\mathcal{F}(\mathcal{D})$)
% \STATE $low \leftarrow 0$; $high \leftarrow n$; $OptScore\leftarrow \infty$
% \WHILE{$low \neq high$}
%     \STATE $mid \leftarrow \frac{low+high}{2}$
%     \STATE $MidScore \leftarrow $\texttt{ComputeScore}$(\mathcal{D}, \tau,\beta =points[mid])$
%     \STATE $NieghborScore \leftarrow $\texttt{ComputeScore}$(\mathcal{D}, \tau,\beta =points[mid+1])$
%     \IF{$MidScore > NieghborScore$} 
%         \STATE $high \leftarrow mid-1$ ; $OptScore\leftarrow NieghborScore$
%     \ELSE
%         \STATE $low \leftarrow mid$ ; $OptScore\leftarrow MidScore$
%     \ENDIF
% \ENDWHILE
\STATE $OptScore,\ LineOpt \leftarrow$\texttt{BinarySearch2D}($\mathcal{D}$, $\tau$, $points$) \COMMENT{$LineOpt$ is the dual line with the least score in plane $z_i=v$}
% \STATE $line_0 \leftarrow $ Get line which corresponds to $points[low]$
% \STATE Compute intersections of the $line_0$ (dual line) with all the other $n-1$ dual lines
% \STATE Let $line_1$ and $line_2$ be the lines which intersect with $line$ to the left and right of where $z_i=v$ intersected with $line$
% \STATE $score_1=$ \texttt{ComputeScore}(\texttt{Intersect}($line_0, line_1$)) 
% \STATE $score_2=$ \texttt{ComputeScore}(\texttt{Intersect}
% ($line_0, line_2$))
% \IF{$score_1 < OptScore$}
%     \STATE $R\leftarrow (R_s, v)$
% \ELSIF{$score_2 < OptScore$}
%     \STATE $R\leftarrow (v, R_e)$
% \ELSE
%     \STATE $R\leftarrow [v, v]$
% \ENDIF
\STATE \textbf{return} \texttt{ComputeInterval}($\mathcal{D}$, $\tau$, $R$, $A'$, $OptLine$, $OptScore$) \COMMENT{Update the interval $R$ based on the neighboring intersections to $OptLine$ around $z_i=v$}
\end{algorithmic}
\end{algorithm}

\begin{algorithm}[H]
\caption{\ConsSampl for three and higher dimensions}
\begin{algorithmic}[1]
\label{alg:sampling-points-multi-dimensional}
\STATE {\bf Input:} Point set $\mathcal{D}$, $\tau$, search interval $R$, variable $A'$, dimension $d$
\STATE {\bf Output:} Sampled vertex $P$
\STATE $Sampled \leftarrow []$ \COMMENT{Empty list}
\FOR{$i=1$ to $d$}
    \STATE Initialize sampling $weights$ to $0$s
    \FOR{$j\in \mathcal{D}$}
        \STATE Create dataset $\mathcal{D}'$ from $\mathcal{D}$ based on the intersections with $sampled$ hyperplanes
        \STATE Update $weights$ based on \ConsSamplFn($\mathcal{D}$, $\tau$, $R$, $A'$, $d-i$)
    \ENDFOR
    \STATE $SampledPlane \leftarrow$ Sample from $\mathcal{D}$ using $weights$
    \STATE $\mathcal{D}\leftarrow\mathcal{D} \setminus SampledPlane$
    \STATE Append $SampledPlane$ into $Sampled$
\ENDFOR
\STATE \textbf{return} \texttt{Intersect}($Sampled$)
\end{algorithmic}
\end{algorithm}

\begin{algorithm}[H]
\caption{\SplitOrc for three and higher dimensions}
\begin{algorithmic}[1]
\label{alg:split-pointers-multi-dimensional}
\STATE {\bf Input:} Point set $\mathcal{D}$, $\tau$, vertex $P_j$ in $S$, search interval $R$, variable $A'$
\STATE {\bf Output:} Updated search interval $R$

\STATE $v \leftarrow P_j[A']$
\STATE Create point set $\mathcal{D}'$ from $\mathcal{D}$ by intersection with hyperplane $z_i=v$
\STATE $OptPoint, OptValue \leftarrow$ \randQR$(\mathcal{D}', \tau)$ 
\STATE $neighbors \leftarrow$ Compute neighbors of $OptPoint$
\FOR{$neighbour in neighbors$}
    \STATE $NeighborVal \leftarrow$ \texttt{ComputeScore}$(\mathcal{D}, \tau, neighbor)$
    \IF{$NeighborVal < OptVal$}
        \STATE \textbf{return} Interval containing $neighbor$
    \ENDIF
\ENDFOR
\STATE \textbf{return} $[v, v]$
\end{algorithmic}
\end{algorithm}

%% file: main.bbl
\begin{thebibliography}{31}
\providecommand{\natexlab}[1]{#1}
\providecommand{\url}[1]{\texttt{#1}}
\expandafter\ifx\csname urlstyle\endcsname\relax
  \providecommand{\doi}[1]{doi: #1}\else
  \providecommand{\doi}{doi: \begingroup \urlstyle{rm}\Url}\fi

\bibitem[Barrodale and Roberts(1973)]{barrodale1973improved}
Ian Barrodale and Frank~DK Roberts.
\newblock An improved algorithm for discrete $l_1$ linear approximation.
\newblock \emph{SIAM Journal on Numerical Analysis}, 10\penalty0 (5):\penalty0
  839--848, 1973.

\bibitem[Bloomfield and Steiger(1983)]{bloomfield1983least}
Peter Bloomfield and William~L Steiger.
\newblock \emph{Least absolute deviations: theory, applications, and
  algorithms}.
\newblock Springer, 1983.

\bibitem[Cade and Noon(2003)]{cade2003gentle}
Brian~S Cade and Barry~R Noon.
\newblock A gentle introduction to quantile regression for ecologists.
\newblock \emph{Frontiers in Ecology and the Environment}, 1\penalty0
  (8):\penalty0 412--420, 2003.

\bibitem[Chan(1999)]{chan1999remarks}
Timothy~M Chan.
\newblock Remarks on k-level algorithms in the plane, 1999.

\bibitem[Chan(2001)]{chan2001dynamic}
Timothy~M Chan.
\newblock Dynamic planar convex hull operations in near-logarithmic amortized
  time.
\newblock \emph{Journal of the ACM (JACM)}, 48\penalty0 (1):\penalty0 1--12,
  2001.

\bibitem[Chernick(2002)]{chernick2002elements}
Michael~R Chernick.
\newblock The elements of statistical learning: Data mining, inference and
  prediction, 2002.

\bibitem[Chernozhukov et~al.(2020)Chernozhukov, Fern{\'a}ndez-Val, and
  Melly]{chernozhukov2020fast}
Victor Chernozhukov, Iv{\'a}n Fern{\'a}ndez-Val, and Blaise Melly.
\newblock Fast algorithms for the quantile regression process.
\newblock \emph{Empirical economics}, pages 1--27, 2020.

\bibitem[Davino et~al.(2013)Davino, Furno, and Vistocco]{davino2013quantile}
Cristina Davino, Marilena Furno, and Domenico Vistocco.
\newblock \emph{Quantile regression: theory and applications}, volume 988.
\newblock John Wiley \& Sons, 2013.

\bibitem[Dey(1997)]{dey1997improved}
Tamal~K Dey.
\newblock Improved bounds on planar k-sets and k-levels.
\newblock In \emph{Proceedings 38th Annual Symposium on Foundations of Computer
  Science}, pages 156--161. IEEE, 1997.

\bibitem[Deza et~al.(2008)Deza, Nematollahi, and Terlaky]{deza2008good}
Antoine Deza, Eissa Nematollahi, and Tam{\'a}s Terlaky.
\newblock How good are interior point methods? klee--minty cubes tighten
  iteration-complexity bounds.
\newblock \emph{Mathematical Programming}, 113\penalty0 (1):\penalty0 1--14,
  2008.

\bibitem[Edelsbrunner(1987)]{edelsbrunner1987algorithms}
Herbert Edelsbrunner.
\newblock \emph{Algorithms in combinatorial geometry}, volume~10.
\newblock Springer Science \& Business Media, 1987.

\bibitem[Edelsbrunner and Welzl(1986)]{edelsbrunner1986constructing}
Herbert Edelsbrunner and Emo Welzl.
\newblock Constructing belts in two-dimensional arrangements with applications.
\newblock \emph{SIAM Journal on Computing}, 15\penalty0 (1):\penalty0 271--284,
  1986.

\bibitem[Edgeworth(1888)]{edgeworth1888xxii}
Francis~Ysidro Edgeworth.
\newblock Xxii. on a new method of reducing observations relating to several
  quantities.
\newblock \emph{The London, Edinburgh, and Dublin Philosophical Magazine and
  Journal of Science}, 25\penalty0 (154):\penalty0 184--191, 1888.

\bibitem[Feng et~al.(2015)Feng, Chen, and He]{feng2015bayesian}
Yang Feng, Yuguo Chen, and Xuming He.
\newblock Bayesian quantile regression with approximate likelihood.
\newblock \emph{Bernoulli}, 21\penalty0 (2):\penalty0 832--850, 2015.

\bibitem[Haupt et~al.(2014)Haupt, L{\"o}sel, and Stemmler]{haupt2014quantile}
Harry Haupt, Friedrich L{\"o}sel, and Mark Stemmler.
\newblock Quantile regression analysis and other alternatives to ordinary least
  squares regression.
\newblock \emph{Methodology}, 2014.

\bibitem[https://cran.r
  project.org/web/packages/quantreg/index.html()]{quantreg}
https://cran.r project.org/web/packages/quantreg/index.html.

\bibitem[Jiang et~al.(2020)Jiang, Song, Weinstein, and Zhang]{jiang2020faster}
Shunhua Jiang, Zhao Song, Omri Weinstein, and Hengjie Zhang.
\newblock Faster dynamic matrix inverse for faster lps.
\newblock \emph{arXiv preprint arXiv:2004.07470}, 2020.

\bibitem[John and Nduka(2009)]{john2009quantile}
Onyedikachi~O John and Ethelbert~C Nduka.
\newblock Quantile regression analysis as a robust alternative to ordinary
  least squares.
\newblock \emph{Scientia Africana}, 8\penalty0 (2):\penalty0 61--65, 2009.

\bibitem[Kleinberg and Tardos(2006)]{kleinberg2006algorithm}
Jon Kleinberg and Eva Tardos.
\newblock \emph{Algorithm design}.
\newblock Pearson Education India, 2006.

\bibitem[Koenker and Bassett~Jr(1978)]{koenker1978regression}
Roger Koenker and Gilbert Bassett~Jr.
\newblock Regression quantiles.
\newblock \emph{Econometrica: journal of the Econometric Society}, pages
  33--50, 1978.

\bibitem[Koenker and d'Orey(1987)]{koenker1987algorithm}
Roger~W Koenker and Vasco d'Orey.
\newblock Algorithm as 229: Computing regression quantiles.
\newblock \emph{Applied statistics}, pages 383--393, 1987.

\bibitem[Meinshausen and Ridgeway(2006)]{meinshausen2006quantile}
Nicolai Meinshausen and Greg Ridgeway.
\newblock Quantile regression forests.
\newblock \emph{Journal of machine learning research}, 7\penalty0 (6), 2006.

\bibitem[Olsen et~al.(2017)Olsen, Tian, Wallace, Nickel, Warren, Fraser,
  Selvam, and Hamilton]{olsen2017use}
Margaret~A Olsen, Fang Tian, Anna~E Wallace, Katelin~B Nickel, David~K Warren,
  Victoria~J Fraser, Nandini Selvam, and Barton~H Hamilton.
\newblock Use of quantile regression to determine the impact on total health
  care costs of surgical site infections following common ambulatory
  procedures.
\newblock \emph{Annals of surgery}, 265\penalty0 (2):\penalty0 331, 2017.

\bibitem[Portnoy and Koenker(1997)]{portnoy1997gaussian}
Stephen Portnoy and Roger Koenker.
\newblock The gaussian hare and the laplacian tortoise: computability of
  squared-error versus absolute-error estimators.
\newblock \emph{Statistical Science}, 12\penalty0 (4):\penalty0 279--300, 1997.

\bibitem[Savva et~al.(2020)Savva, Anagnostopoulos, and
  Triantafillou]{savva2020ml}
Fotis Savva, Christos Anagnostopoulos, and Peter Triantafillou.
\newblock Ml-aqp: Query-driven approximate query processing based on machine
  learning.
\newblock \emph{arXiv preprint arXiv:2003.06613}, 2020.

\bibitem[Thirumuruganathan et~al.(2022)Thirumuruganathan, Shetiya, Koudas, and
  Das]{thirumuruganathan2022prediction}
Saravanan Thirumuruganathan, Suraj Shetiya, Nick Koudas, and Gautam Das.
\newblock Prediction intervals for learned cardinality estimation: An
  experimental evaluation.
\newblock In \emph{2022 IEEE 38th International Conference on Data Engineering
  (ICDE)}, pages 3051--3064. IEEE, 2022.

\bibitem[Vaidya(1989)]{vaidya1989speeding}
Pravin~M Vaidya.
\newblock Speeding-up linear programming using fast matrix multiplication.
\newblock In \emph{30th annual symposium on foundations of computer science},
  pages 332--337. IEEE Computer Society, 1989.

\bibitem[Wagner(1959)]{wagner1959linear}
Harvey~M Wagner.
\newblock Linear programming techniques for regression analysis.
\newblock \emph{Journal of the American Statistical Association}, 54\penalty0
  (285):\penalty0 206--212, 1959.

\bibitem[Wright(1997)]{wright1997primal}
Stephen~J Wright.
\newblock \emph{Primal-dual interior-point methods}.
\newblock SIAM, 1997.

\bibitem[Yang et~al.(2013)Yang, Meng, and Mahoney]{yang2013quantile}
Jiyan Yang, Xiangrui Meng, and Michael Mahoney.
\newblock Quantile regression for large-scale applications.
\newblock In \emph{International Conference on Machine Learning}, pages
  881--887. PMLR, 2013.

\bibitem[Zheng(2011)]{zheng2011gradient}
Songfeng Zheng.
\newblock Gradient descent algorithms for quantile regression with smooth
  approximation.
\newblock \emph{International Journal of Machine Learning and Cybernetics},
  2\penalty0 (3):\penalty0 191--207, 2011.

\end{thebibliography}
